# Wafer-scale Integration of Single-Crystalline $MoS_2$ for Flexible Electronics Enabled by Oxide Dry-transfer

Xiang Xu[1,2†], Yitong Chen[1,2†], Jichuang Shen[1,2†], Qi Huang[4], Tong Jiang[1,2], Han Chen[1,2], Huaze Zhu[2], Yaqing Ma[1,2], Hao Wang[2], Wenhao Li[1,2], Chen Ji[2], Dingwei Li[1,2], Siyu Zhang[2], Yan Wang[5], Bowen Zhu[2,3,4,*], Wei Kong[2,3,4,*]

[1] School of Materials Science and Engineering, Zhejiang University, Hangzhou 310027, China.

[2] Key Laboratory of 3D Micro/Nano Fabrication and Characterization of Zhejiang Province, School of Engineering, Westlake University, Hangzhou 310030, China.

[3] Research Center for Industries of the Future, Westlake University, Hangzhou 310030, Zhejiang, China.

[4] Westlake Institute for Optoelectronics, Hangzhou 311421, China.

[5] School of Materials and Energy, University of Electronic Science and Technology of China, Chengdu 610054, China.

[†] These authors contributed equally to this work.

**Abstract**

Atomically thin, single-crystalline transition metal dichalcogenides (TMDCs) grown via chemical vapor deposition (CVD) on sapphire substrates exhibit exceptional mechanical and electrical properties, positioning them as excellent channel materials for flexible electronics. However, conventional wet-transfer processes for integrating these materials onto flexible substrates often introduce surface contamination, significantly degrading device performance. Here, we present a wafer-scale dry-transfer technique using a high-κ dielectric oxide as the transfer medium, enabling the integration of 4-inch single-crystalline $MoS_2$ onto flexible substrates. This method eliminates contact with polymers or solvents, thus preserving the intrinsic electronic properties of $MoS_2$. As a result, the fabricated flexible field-effect transistor (FET) arrays exhibit remarkable performance, with a mobility of 117 $cm^2/V \cdot s$, a subthreshold swing of 68.8 mV $dec^{-1}$, and an ultra-high current on/off ratio of $10^{12}$—values comparable to those achieved on rigid substrates. Leveraging the outstanding electrical characteristics, we demonstrated $MoS_2$-based flexible inverters operating in the subthreshold regime, achieving both a high gain of 218 and ultra-low power consumption of 1.4 pW/μm. Additionally, we integrated a flexible tactile sensing system driven by active-matrix $MoS_2$ FET arrays onto a robotic gripper, enabling real-time object identification. These findings demonstrate the simultaneous achievement of high electrical performance and flexibility, highlighting the immense potential of single-crystalline TMDC-based flexible electronics for real-world applications.

## Introduction

Flexible electronics represent a rapidly advancing field with vast potential in applications such as wearable technology, robotics, and health monitoring systems[1-7]. A crucial element in achieving high-performance flexible devices is the development of flexible channel materials, which must combine mechanical flexibility with excellent electronic properties[8-11]. Among the materials explored, transition metal dichalcogenides (TMDCs) stand out for their exceptional mobility and mechanical robustness in atomically thin structures, offering flexibility without compromising electronic performance—making them prime candidates for flexible electronics[12-15].

Recent advancements have enabled the epitaxial growth of single-crystalline TMDCs on sapphire substrates using chemical vapor deposition (CVD) techniques[16-19]. These single-crystalline TMDCs outperform their polycrystalline counterparts, which are often hindered by defects and grain boundaries[20, 21]. These imperfections typically introduce electron scattering and leakage, reducing both mobility and device reliability, thereby compromising overall device performance[22-24].

Despite these advantages, integrating wafer-scale single-crystalline TMDCs into flexible platforms presents a significant challenge due to their typical growth on rigid sapphire substrates[25-28]. The strong adhesion between the single-crystalline TMDC layer and the sapphire substrate complicates the process of transferring the material onto flexible substrates[29-31]. Current transfer methods often rely on wet-transfer approaches, which involve solvents or chemical agents to intercalate or etch the TMDC-sapphire interface, releasing the material from the substrates[32-35]. Polymers like polymethyl methacrylate (PMMA) are frequently employed to facilitate this transfer onto flexible substrates. However, the use of polymers and solvents in direct contact with the TMDC can degrade the material's intrinsic electronic properties, leading to suboptimal device performance[36, 37]. For instance, polymer residues can create charge traps, resulting in mobility reduction and doping variation[38, 39]. Furthermore, these chemical processes are not fully compatible with standard microfabrication techniques, posing challenges for scalability and repeatability in mass production[40-42].

In this study, we introduce a wafer-scale dry-transfer technique utilizing a high-κ dielectric oxide as the transfer medium, enabling the integration of wafer-scale single-crystalline $MoS_2$ onto flexible substrates. This approach is fully compatible with standard microfabrication workflows, avoiding contact with polymers or solvents, thus preserving the intrinsic electronic properties of $MoS_2$. Consequently, the fabricated flexible field-effect transistor (FET) arrays exhibit remarkable performance and wafer-scale uniformity. Building on these properties, we demonstrate $MoS_2$-based flexible inverters capable of operating in the subthreshold regime, achieving record low power consumption at picowatt level. Additionally, we developed a flexible tactile sensing system powered by active-matrix $MoS_2$ FET arrays, conformally integrated onto a robotic gripper for real-time object identification. These results highlight the significant potential of single-crystalline $MoS_2$ for high performance flexible electronics in practical applications.

**Oxide dry-transfer of single-crystalline TMDCs to flexible substrates**

Figure 1a illustrates the oxide dry-transfer process for single-crystalline transition metal dichalcogenides (TMDCs), in this case using $MoS_2$ grown on sapphire. Monolayer, single-crystalline $MoS_2$ was epitaxially synthesized on c-plane sapphire via chemical vapor deposition (CVD) (see Methods), yielding high-quality crystallinity (see Supplementary Figures 1-3). A thin layer of $Al_2O_3$ was first deposited onto the $MoS_2$ surface using electron beam (E-beam) evaporation for adhesion enhancement, followed by atomic layer deposition (ALD) of $Al_2O_3$ as the high-κ dielectric. Subsequently, a metal layer was deposited atop the oxide, and the entire structure was adhered to a flexible substrate, such as polyethylene terephthalate (PET), and is exfoliated from the sapphire substrate.

The exfoliation process resulted in a wafer-scale $MoS_2$/high-κ $Al_2O_3$/metal multilayer structure on the flexible PET substrate (see Fig. 1b). Importantly, throughout the transfer, the surface of the TMDC avoids contacting with water or polymers, ensuring the preservation of the intrinsic properties of these high-quality single-crystalline materials. Low-energy electron diffraction (LEED) analysis (Fig. 1c) confirmed that the single-crystallinity of $MoS_2$ was retained after transfer to the flexible substrate. This oxide dry-transfer process is compatible with other CVD-grown single-crystalline TMDCs on sapphire, such as $WS_2$ and $WSe_2$ (see Supplementary Fig. 4).

For device array fabrication, the metal layer on the high-κ dielectric can be photolithographically patterned into gate arrays prior to attachment to the flexible substrate, enabling the formation of local back gates for individual device control. After the transfer, the top surface of the wafer-scale TMDC is exposed, allowing for contact fabrication and potential top-gating for dual-gate control (see Fig. 1d, Supplementary Fig. 5 and Note 1).

**Material quality of oxide dry-transferred single-crystalline $MoS_2$**

The quality of the oxide dry-transfer process was rigorously evaluated. Optical microscopy images revealed a uniform surface with no observable contamination across the entire sample of oxide dry-transferred $MoS_2$, maintaining surface integrity comparable to that of as-grown $MoS_2$ on sapphire (Supplementary Fig. 6). Scanning electron microscopy (SEM) and atomic force microscopy (AFM) images further confirmed that the material was free from wrinkles, cracks, or residuals at the microscopic level. In contrast, such defects and contamination were evident in conventional wet-transfer processes, both macroscopically and microscopically (Fig. 2a-b and Supplementary Fig. 7). X-ray photoelectron spectroscopy (XPS) provided additional insights into the surface chemistry of each approach (Fig. 2c). Prominent C-O and C-H signals were observed in wet-transferred samples, corresponding to residuals of PMMA and organic solvents, the contamination of which is absent on the surface of oxide dry-transferred and as-grown $MoS_2$.

Due to the absence of dangling bonds on the van der Waals surface of two-dimensional materials, thin-film atomic layer deposition (ALD) of oxides on TMDC materials faces significant challenges in nucleation and convergence leading to poor coverage and dielectric quality reduction[43, 44]. Additionally, the bonding between TMDC materials and ALD oxides is generally weaker compared to that between single-crystalline TMDC materials and their substrates, which can result in partial delamination during TMDC material exfoliation (see Supplementary Fig. 8 and Note 2). To overcome these challenges, we first deposited an ultra-thin oxide layer on $MoS_2$ (Supplementary Fig. 9) using electron beam (e-beam) evaporation. As physical vapor deposition, the accumulation of relatively immobile deposited oxide species allows the formation of a dense, uniform layer on the van der Waals surface, improving adhesion to the $MoS_2$ compared to ALD oxides. This enhanced adhesion is evidenced by the blue shift of the XPS S 2p peak after transfer (Supplementary Fig. 10), indicating a stronger binding energy between $MoS_2$ and the e-beam-deposited $Al_2O_3$.

Subsequently, we deposited an additional layer of ALD $Al_2O_3$ (Supplementary Fig. 11) on the fully converged e-beam oxide surface to achieve superior dielectric properties. Capacitance tests (Fig. 2i) confirmed high-quality dielectric performance of the composite oxide layer, with a dielectric constant over 7, approaching the ideal dielectric constant of ALD-prepared $Al_2O_3$ (κ=7-9)[45, 46], thereby enabling efficient gate control. Cross-sectional transmission electron microscopy (TEM) further confirmed that the oxide layer deposition process did not induce damage or defects in the single-crystalline $MoS_2$ (Fig. 2g and Supplementary Fig. 12). Additionally, the near-identical XPS Mo-3d spectra, as well as Raman and photoluminescence (PL) measurements of $MoS_2$ before and after transfer (Fig. 2d-f), indicate that the transfer process had minimal impact on the intrinsic properties of $MoS_2$.

**Electrical characteristics of oxide dry-transferred single-crystalline $MoS_2$ FETs**

The ultra-clean oxide dry-transfer technique facilitated the fabrication of high-performance, wafer-scale flexible device arrays. Specifically, an array of 432 × 432 back-gated field effect transistors (FETs) was produced in 4-inch diameter. The channel material consisted of monolayer single-crystalline $MoS_2$, with dimensions of 8 μm in length and 20 μm in width, as shown in Figure 3a-c (refer to Methods, Supplementary Figure 13 and Note 3 for fabrication process details).

The output and transfer characteristics of the oxide dry-transferred devices demonstrated significant advantages over those of wet-transferred devices (Figures 2d-e). The off-state current of the oxide dry-transferred FETs was measured at approximately $10^{-15}$ A, which is four orders of magnitude lower than the $10^{-11}$ A observed in wet-transferred devices, with the off-state current approaching the femtoampere level, close to the measurement limit. Additionally, the sub-threshold swing (SS) was measured at 68.8 mV $dec^{-1}$, approaching the theoretical minimum for thermionic emission devices, compared to 216 mV $dec^{-1}$ for wet-transferred devices. This marked improvement is attributed to the oxide dry-transfer technique, which significantly reduces impurity carriers and interface state density ($D_{it}$) in the channel material. The $D_{it}$, extracted using the subthreshold slope method, was approximately $9.6 \times 10^{10}$ $cm^{-2}$ $eV^{-1}$ for the oxide dry-transferred devices—two orders of magnitude lower than the $1.8 \times 10^{12}$ $cm^{-2}$ $eV^{-1}$ observed in wet-transferred devices. This reduction in $D_{it}$ enables faster switching and ultra-low off-state currents, demonstrating the superior gate control achieved by the oxide dry-transferred FETs.

Moreover, the on-state current of the oxide dry-transferred devices reached approximately 300 μA, compared to 100 μA in wet-transferred devices. This enhancement is attributed to the absence of solvent residues in the oxide dry-transfer process, which improves metal-semiconductor contact. Figure 3f shows the contact resistance ($R_c$) of the oxide dry-transferred devices was measured at 214 Ω·μm using Transmission Line Method (TLM) (see Supplementary Fig. 14), a fourfold reduction compared to the 759 Ω·μm of wet-transferred devices. The combination of high on-state current and low off-state current results in an ultra-high on/off ratio close to $10^{12}$, placing these devices among the highest-performing $MoS_2$-based devices. Additionally, a maximum mobility of 117 $cm^2/V\cdot s$ was achieved, compared to 52 $cm^2/V\cdot s$ for wet-transferred devices (refer to Supplementary Note 4 for calculation details). Statistical analysis, as shown in Supplementary Figure 15 and Figure 4g, confirms the uniform performance across the 4-inch scale device arrays, highlighting the scalability and robustness of the oxide dry-transfer technique.

In comparison to previously reported $MoS_2$ FETs, our oxide dry-transferred flexible devices exhibit performance on par with the best-reported devices on rigid substrates[47], a significant improvement over earlier flexible $MoS_2$ devices using poly-crystalline materials and wet-transfer method (Fig. 3h)[48, 49]. Notably, the high performance of the $MoS_2$ device arrays was maintained during bending tests, withstanding mechanical strain with a radius of curvature as small as 5 mm (Figure 3i and Supplementary Figure 16). Figure 3j compares the performance of commonly used large-area flexible thin-film transistors (TFTs), such as amorphous silicon (a-Si), low-temperature polycrystalline silicon (LTPS), oxide materials (IGZO, ZnO, etc.), and organic materials[50]. Our oxide dry-transferred flexible devices demonstrate both low leakage currents and high mobility, outperforming conventional flexible material systems. These results indicate that our devices hold strong potential for applications in the most requiring scenarios, where both high processing capability and flexibility are critical.

**Ultra-low power single-crystalline $MoS_2$ flexible logic inverters**

The local back-gating configuration, enabled by the oxide dry-transfer approach, allows device interconnections and integration into logic circuits (see Supplementary Figure 5 and Note 2). By leveraging the steep subthreshold swing and low off-state current of our single-crystalline $MoS_2$ field-effect transistors (FETs), we fabricated inverter arrays on a flexible PET substrate at wafer scale (see Figure 4a-c). The nMOS inverter was constructed with one load and one drive FET. We initially investigated the intrinsic gain ($A_i = g_m \times r_o$) of a single transistor, which depends on the transconductance ($g_m$) and output resistance ($r_o$). As shown in Figure 4d, the extracted intrinsic gain $A_i$ ranged from $10^3$ to $10^4$ within the $V_{gs}$ range of 0-1 V, comparable to the highest values achieved in 2D materials and more than two orders of magnitude higher than that of silicon[51]. This impressive gain is attributed to the high $g_m$ and large $r_o$ in the deep subthreshold region, corresponding to the steep subthreshold swing and nearly zero increase in current during output saturation (see Supplementary Figure 17)[52-54]. Consequently, the voltage transfer characteristics (VTC) of the inverter exhibit sharp output voltage transitions (see Figure 4e). The inverter achieved a peak voltage gain of 216 at $V_{dd}$ = 5 V (see Figure 4f). We also conducted bending tests on the flexible inverter, which showed stable logical operation under mechanical strain, with negligible performance degradation, demonstrating the robustness of these flexible transistors (see Figure 4g).

Additionally, we evaluated the inverter's power consumption ($P = I_{dd} \times V_{dd}$), which is critical for low-power, high-performance logic computing and signal amplification[55, 56]. The inverter showed exceptionally low power consumption, ranging from approximately 7.7 to 29 pW under a $V_{dd}$ of 1-5 V at the transition voltage (see Figure 4h). The operating transition point was around -0.35 V, situated in the deep-subthreshold region. Due to the ultra-low off-state current of the FET, the inverter operates with minimal current in this region. As a result, our inverter exhibits the lowest power consumption reported among all 2D material-based systems[43, 57]. To benchmark its overall performance across various material systems and structures, we compared the gain-power consumption values of our oxide dry-transfer $MoS_2$-based inverter with other reported thin-film inverters on flexible substrates (see Figure 4i). The power consumption level has reached the pW level (1.4 pW/μm), which is the lowest power consumption level compared to other thin-film flexible inverters. Additionally, compared to inverters that have achieved power consumption in the pW range, such as those based on organic materials and CNTs, our device offers more than double the gain. These findings demonstrate the significant potential of single-crystalline $MoS_2$ for low-power, high-performance flexible circuit applications.

**Integration of flexible $MoS_2$ FET arrays for active-matrix tactile sensing**

In robotic tactile sensing, large-scale integrated pressure sensor arrays with enhanced spatial resolution are crucial for accurately detecting the location, shape, size, and pressure distribution during interactions between the robot and its environment[58, 59]. In this work, we developed an active-matrix tactile sensor (AM-TS) array based on wafer-scale single-crystalline $MoS_2$, consisting of a 10×10 array of sensor units (see Figure 5a, b and Methods). Each sensor unit comprises a $MoS_2$ transistor paired with a piezoresistive sensor. The piezoresistive sensor, featuring pyramid microstructures made of CNTs/PDMS (see Methods), exhibits high sensitivity to low-pressure conditions (see Supplementary Figure 18 and Note 5). The sensor units are organized in an active-matrix configuration (see Figure 5c), where each $MoS_2$ transistor functions as a switch controlling the piezoresistive sensor connected in series. This configuration ensures that only the addressed sensor element transmits data, thereby providing high-sensitivity pressure sensing with real-time spatial resolution.

The mechanical flexibility of the $MoS_2$ field-effect transistors (FETs) allows them to conform to the shape of the sensor array without performance degradation, enabling the fabrication of flexible and conformal AM-TS devices. The I-V characteristics of the 10×10 $MoS_2$ transistor array in the AM-TS sensor demonstrate consistent switching performance (see Figure 5d). Notably, the drain current of the access transistor increases discernibly in response to rising pressure, with sensitivity sufficient to detect pressure changes as low as 200 Pa (see Figure 5e-f). This high sensitivity enables precise pressure mapping, even at low-pressure thresholds, allowing accurate tactile sensing.

Figure 5g demonstrates the integration of a single-crystalline $MoS_2$-based active-matrix tactile sensor (AM-TS) with a soft robotic gripper, where the flexible AM-TS is conformally attached to the gripper's finger for precise object-handling tasks. The substantial drain current response allows for accurate pressure mapping, enabling the identification of object shapes, locations, and sizes with precision. Figures 5h and 5i show the signals recorded by the AM-TS in response to pressure stimuli, which effectively translate pressure distribution into corresponding current changes, successfully detecting the shape of a nut. The real-time pressure mapping capability, enabled by the flexible single-crystalline $MoS_2$ field-effect transistors (FETs), allows the high sensitivity and spatial resolution of tactile sensing in a compact design. This feature is particularly valuable for applications such as robust object manipulation, precise placement, and object classification in cluttered environments, where simultaneous detection, handling, and navigation are essential.

## Conclusion

In conclusion, we have introduced an oxide-based dry transfer technique for integrating monolayer single-crystalline $MoS_2$ onto flexible substrates, eliminating the use of polymers and water while maintaining compatibility with standard microfabrication processes. This method resulted in an enhanced mobility of 117 $cm^2/V{\cdot}s$ and a high on/off current ratio of up to $10^{12}$, representing a significant improvement over the 45 $cm^2/V{\cdot}s$ mobility and $10^7$ on/off ratio achieved using wet transfer techniques. With a subthreshold slope of 68.8 mV/dec, approaching the theoretical limit, we also demonstrated a logic inverter with a high gain of 218 and record-low power consumption at the picowatt level. The performance of flexible $MoS_2$-based devices fabricated using our method is comparable to those on rigid substrates and surpasses those produced with conventional flexible channel materials. Furthermore, we successfully integrated a flexible pressure sensor circuit powered by $MoS_2$ transistors, enabling real-time object identification in a robotic gripper system. These results demonstrate the potential of wafer-scale single-crystalline TMDCs for high-performance flexible electronics in practical applications.

## Methods

### Epitaxial growth of $MoS_2$.

Inch-scale single-crystalline $MoS_2$ are grown on C/A sapphire substrate with a miscut of 1.0° by chemical vapor deposition according to Ref.17. In briefly, the growth of $MoS_2$ was performed in a low-pressure CVD system (~300 Pa) with a tube diameter of 80 mm. S powders (15 g) were placed in a corundum crucible and heated independently at 180°C in zone I and carried by 200 sccm Ar. $MoO_3$ powders (80mg) were heated at 550°C in zone II and carried by 3 sccm $O_2$ and 100 sccm Ar separately from S vapor. The typical substrate temperature in zone III was 950°C and the growth time was 40min.

### Transfer of wafer-scale $MoS_2$.

The oxide dry-transfer method is shown in Figure 1 in the main text and Supplementary Figure 5 and Note 1. Traditional PMMA wet transfer: First, 5% 950 polymethyl methacrylate (PMMA) in anisole was spin-coated on $MoS_2$/sapphire at a speed of 2,000

r.p.m. then baked at 180 °C for 1 min. This process was repeated three times. Second, as-made PMMA/$MoS_2$/sapphire samples were immersed in KOH solution at 110 °C for half an hour. The samples were then picked up from the solution and transferred into deionized water to release the PMMA/$MoS_2$ layer from the sapphire. After full release, the PMMA/$MoS_2$ layers floated to the surface of the deionized water and could be picked up and transferred to targeting substrates. Finally, transferred samples were kept at room temperature for ~12 h to remove the extra water. The PMMA was washed out in acetone.

**Materials characterizations.**

Optical images were taken by Olympus BX51M microscope. The surfaces of 2D materials and oxide thin films were characterized by AFM (Dimension Icon, Bruker). Photoluminescence (PL)/Raman spectra and mapping were measured on alpha 300R (WITec GMBH, Germany) confocal Raman system. A diode-pumped solid-state laser (532 nm, cobalt Laser) was focused on samples with a diffraction-limited beam size of 590 nm by a 50x long-focus objective (NA = 0.55).

XPS measurements were carried out in a PHI5000 system equipped with a monochromatic Al-Kα radiation of 1486.7 eV. All XPS spectra were calibrated using the C 1s peak at 284.8 eV and analyzed via a CasaXPS software.

LEED were performed on an ARPES system connected to an STM and an XPS through an ultrahigh vacuum interconnect system, located at Westlake University. The system was equipped with a DA30-L analyser and a Helium ultrahigh-flux VUV source with a photon energy of 21.2 eV. All tests were performed at T = 5 K, with a base pressure better than $1\times10^{-10}$ mbar. A pass energy of 5 eV was used in all measurements and the analyser slit (with aperture) was set to 0.5 mm.

SAED and HRTEM were performed using a Thermo-Fisher Talos F200X G2 instrument operating at 200 kV, and EDS mappings were conducted by a Thermo-Fisher Super X. HAADF-STEM images were acquired using a spherical aberration-corrected transmission electron microscope (Thermo Fisher Scientific Spectra Ultra) operating at 300 kV. The probe convergence semi-angle was 21 mrad, and the annular detector collection angle ranges is 69~200 mrad. The cross-sectional TEM samples were prepared by FIB with a Ga+ based focused ion-beam system (Helios G5 UX, Thermo-Fisher). A 15nm carbon and 15nm platinum were first deposited on sample by thermal evaporation (EM ACE600, LEICA), followed by carbon electron beam deposition (200 nm) and platinum ion beam deposition (2μm) to prevent damage and heating effects during focused ion-beam milling. Electrical contacts to two-dimensional semiconductors.

**Fabrication of Wet-transfer $MoS_2$ devices and electrical measurements.**

The oxide dry-transfer method is shown in Supplementary Figure 13 and Note 3. For wet-transfer FETs fabrication, monolayer $MoS_2$ was transferred onto a 35-nm $Al_2O_3$/highly doped Si substrate using the PMMA transfer method. Next, the channel areas were isolated by lithography (Suss-MA/BA6 Gen4, Photoresist: ARP-5350) and $O_2$ dry etching (Samco-RIE-230ip). After etching, the photoresist mask layer on the surface of the channel material is removed by soaking in acetone, isopropanol and deionized water respectively. Then, lithography was used to define the source/drain contacts, followed by e-beam evaporation of 20 nm bismuth/30 nm Au and lift-off. No annealing was performed on the devices. All electrical measurements were carried out by a Keithley 4200A-SCS semiconductor parameter analyzer in a closed-cycle cryogenic probe station with base pressure $10^{-6}$ mba at room temperature.

**Fabrication of the piezoresistive sensor film.**

The base Polydimethylsiloxane (PDMS) elastomer and carbon nanotubes (CNTs) were thoroughly mixed for three hours. Next, the PDMS curing agent (Sylgard 184, Tow Corning) was added to the solution at a weight ratio of 10:1 (PDMS elastomer to PDMS curing agent), followed by an additional 30 minutes of stirring. The resulting mixture was then cast onto a silicon mold with a recessed pyramidal array surface and subjected to expedite the degassing in a vacuum desiccator, followed by heating at 80 °C for one hour. Finally, the fabricated piezoresistive sensor film, incorporating PDMS and CNTs, was cut into squares measuring 1.5 × 1.5 cm for further application.

**Fabrication of the active-matrix tactile sensing array.**

First, the gate electrode (Cu) was patterned on a polyethene terephthalate (PET) substrate. Subsequently, the gate dielectric layer $Al_2O_3$ was deposited, followed by the utilization of large-area single-layer $MoS_2$, transferred via oxide dry-transfer technology, as the active channel layer. Source/drain (S/D) electrodes for the transistor array and the interdigitated electrodes for the sensing area were then deposited. A spin-coating and patterned SU-8 passivation layer was applied on the top of the transistor array to mitigate environmental influence and minimize crosstalk[60]. Finally, the piezoresistive sensor film was laminated onto the active-matrix backplane of the 10×10 FETs array to form the AM-TS.

## Author Contribution

X.X., Y.C., and J.S. contributed equally to this work. W.K., B.Z. and X.X. designed the experiment. X.X. performed TMDCs transfer, device fabrication and electrical testing. Y.C. performed the integration and imaging testing of AM-TS. J.S. tested the Raman/PL, AFM. Q.H and D.L participated in the testing and analysis of the inverter. T.J performed analysis of FETs. H.C. and H.W. tested the XPS. H.Z. performed the TEM measurements and FIB. Y.M. performed SEM. W.L provided assistance for the wet-transfer method. J.S. and C.J. performed the CVD growths. S.Z. prepared the pressure sensor. Y.W. provided flexible substrate and test mold. W.K. supervised the project. All authors verified the manuscript.

## Acknowledgments

This work is supported by the National Natural Science Foundation of China (Grant No. 62174138), and the Key Project of Westlake Institute for Optoelectronics (Grant No. 2023GD004). We thank Westlake Center for Micro/Nano Fabrication, the Instrumentation and Service Center for Physical Sciences (ISCPS), and the Instrumentation and Service Center for Molecular Sciences (ISCMS) at Westlake University for the facility support and technical assistance.

## Competing interests

The authors declare no competing interests.

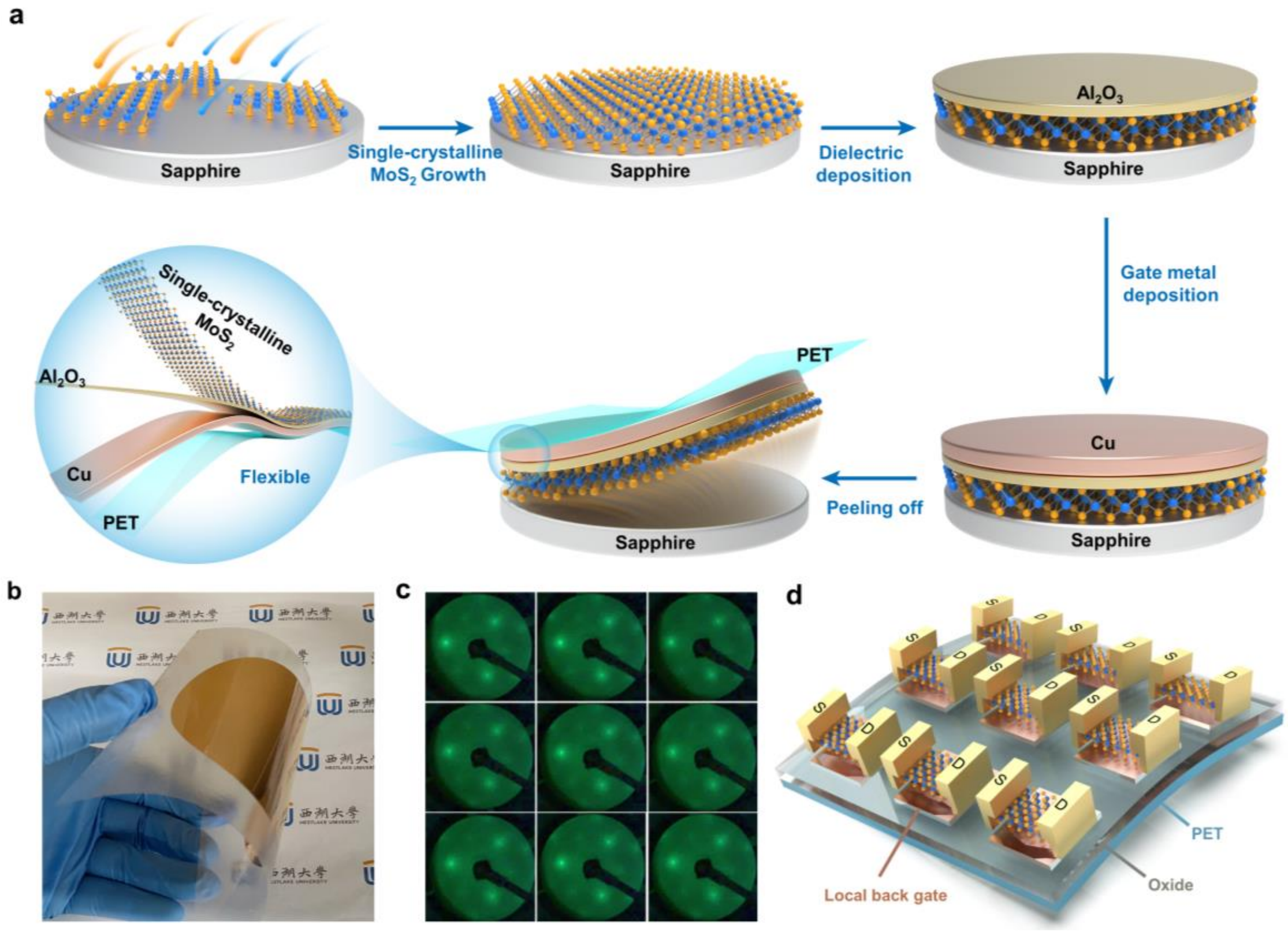

**Fig. 1 Oxide Dry-Transfer Process. a**, Schematic diagram of the $Al_2O_3$ assisted dry-transfer of wafer scale monolayer single-crystalline $MoS_2$. **b**, Optical image of 4-inch single-crystalline $MoS_2$ after transfer. **c**, LEED patterns of $MoS_2$ after transfer scanned across a 1-$cm^2$ sample area with a step of 3 mm. d, Schematic view of local back-gated FETs.

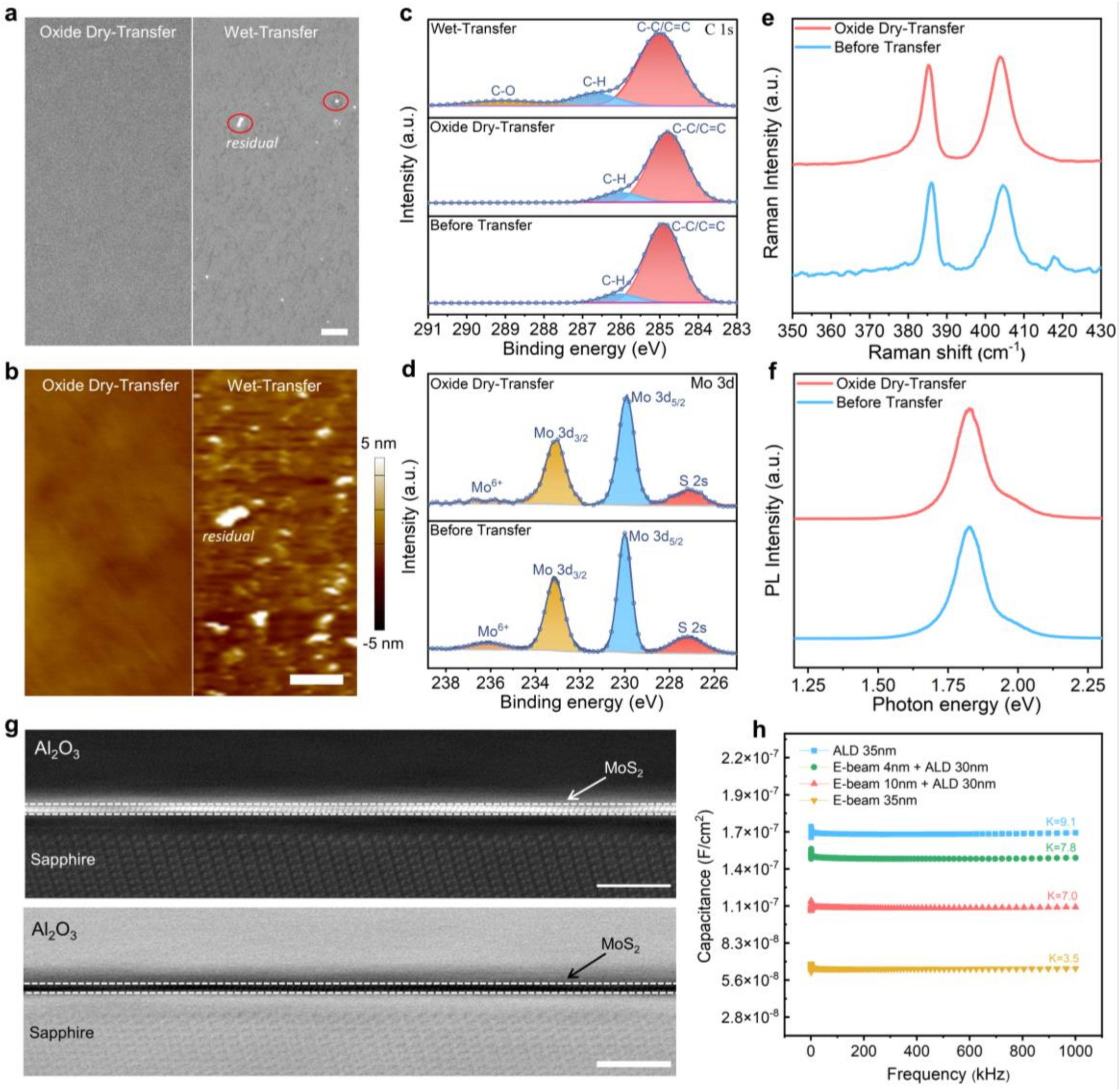

**Fig.2 Material quality analysis of the oxide dry-transfer processes. a-b**, SEM and AFM comparison of transferred $MoS_2$, respectively. Left: oxide dry-transfer. Right: wet-Transfer. Scale bar in SEM, 300 nm. Scale bar in AFM, 2 µm. **c**, C 1s XPS of pristine CVD-$MoS_2$, oxide dry-transfer and wet-transferred $MoS_2$. **d,** Mo 3d XPS of $MoS_2$ before and after oxide dry-transfer. **e-f**, Raman and PL spectrum of monolayer $MoS_2$ before and after oxide dry-transfer. **g,** Cross-sectional HAADF-STEM images in brightfield and darkfield of the $Al_2O_3$/$MoS_2$/Sapphire interface Scale bars, 2 nm. **h**, Comparison of dielectric properties of different dielectric layers.

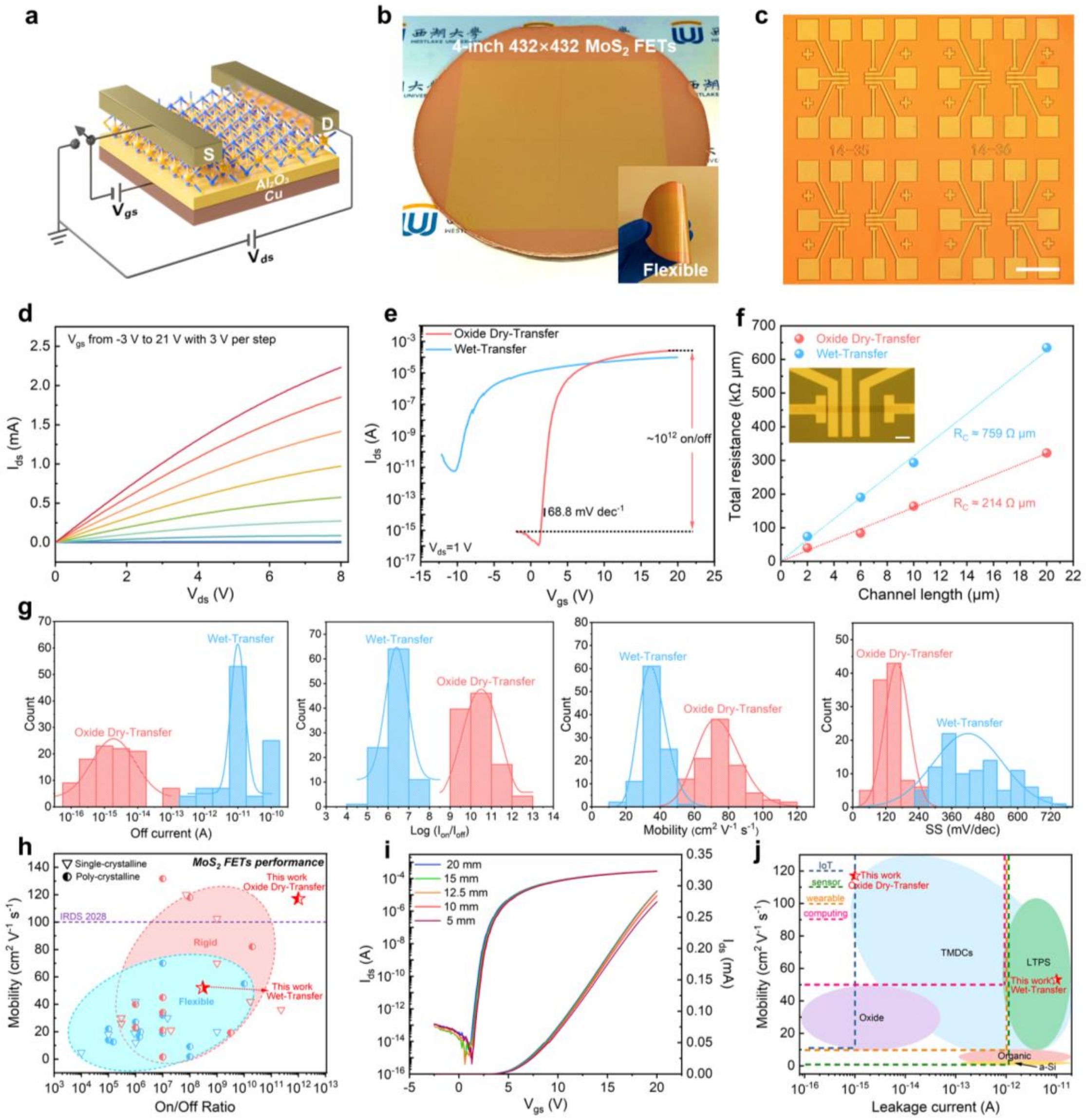

**Fig.3 Performance characterization of the $MoS_2$ transistors. a,** Schematic view of back-gated $MoS_2$ FET. **b,** Photograph of 4-inch wafer-scale flexible $MoS_2$ transistor arrays. Inset: flexibility demonstration. **c,** Microscope image of a fabricated FET array. Scale bar, 200 μm. **d,** Output characteristics of oxide dry-transferred $MoS_2$ FET with channel width and length of 20 and 8 μm. **e,** Comparison of transfer curves of oxide dry-transferred and wet-transferred $MoS_2$ FETs with the same geometry. **f,** TLM measurements of the contact resistance from the wet-transfer and oxide dry-transfer. **g,** Performance comparison of wet-transferred and oxide dry-transferred $MoS_2$ FETs: off current statistics, statistical distribution of on/off ratio, field-effect mobility, subthreshold swing. **h,** Benchmark of the on/off ratio and mobility for flexible/rigid substrate and single/poly-crystalline transistors reported in the literature. The reference data points are listed in Supplementary Table 1 and Table 2. **i,** Transfer curves of flexible $MoS_2$ transistors at different radius of curvature. j, Comparison of performances of the main large-area flexible TFT technologies, the dotted lines with different color represent the requirements of the corresponding TFT application technology.

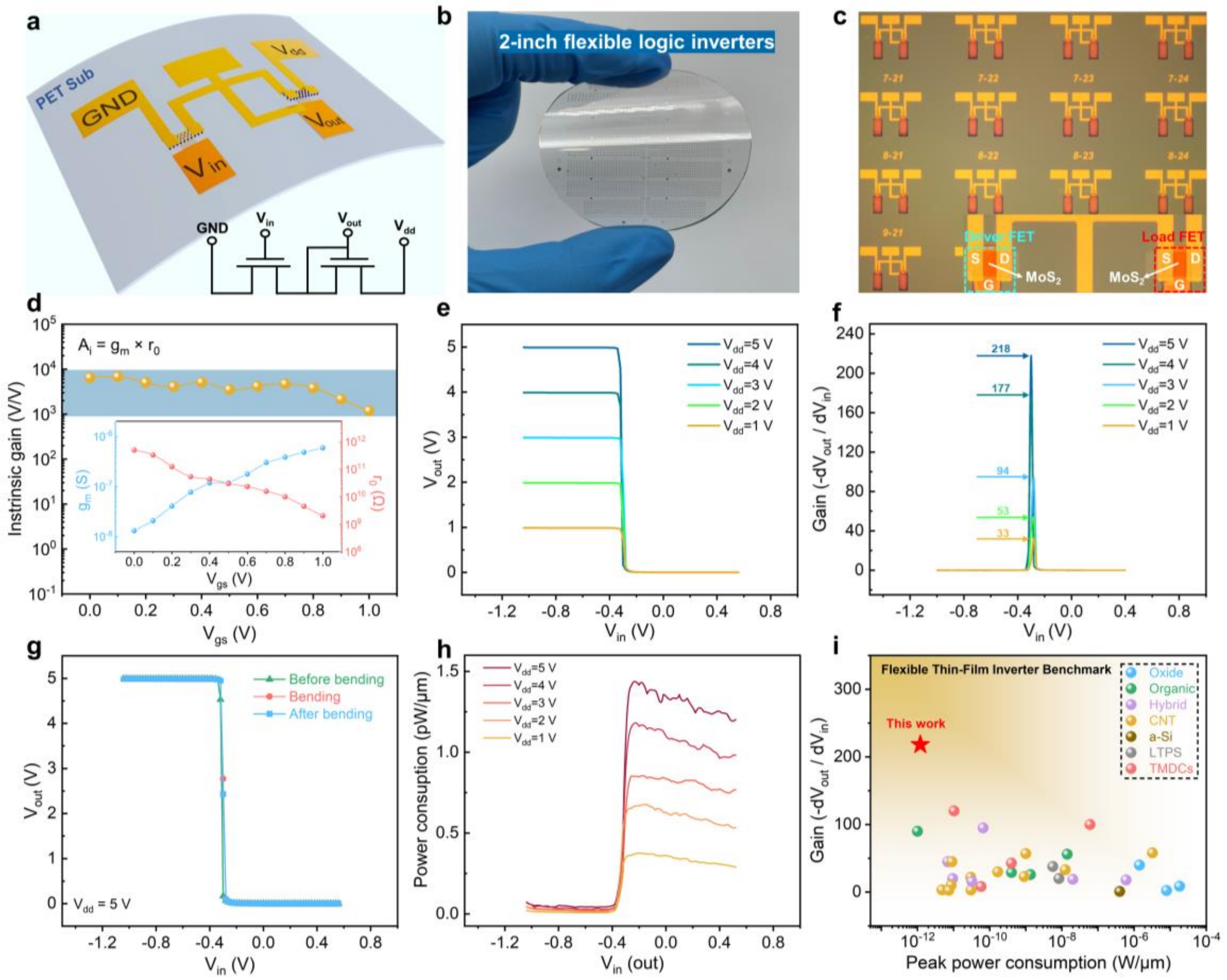

**Fig.4 The Characteristics of the flexible logic $MoS_2$-based inverters. a,** Schematic 3D illustration of $MoS_2$-based inverter built on a PET substrate. Insert: the circuit diagram of inverters. **b,** Photographic image of 2-inch wafer-scale flexible $MoS_2$-based inverter arrays. **c,** Microscope image of a fabricated $MoS_2$-based inverter arrays. **d,** Intrinsic gain as a function of $V_{gs}$. Insert: transconductance ($g_m$) and output resistance ($r_0$) **e,** Voltage transfer characteristics of the inverter at different $V_{dd}$. **f,** Voltage gain of the inverter at different supply voltages. **g,** Voltage transfer characteristics of an inverter under different bending states. **h,** Dynamic power consumption during the scanning process of input voltage. **i,** Summary of the voltage gain and power consumption for various thin-film inverters on flexible substrates reported in the literature. The reference data points are listed in Supplementary Table 3.

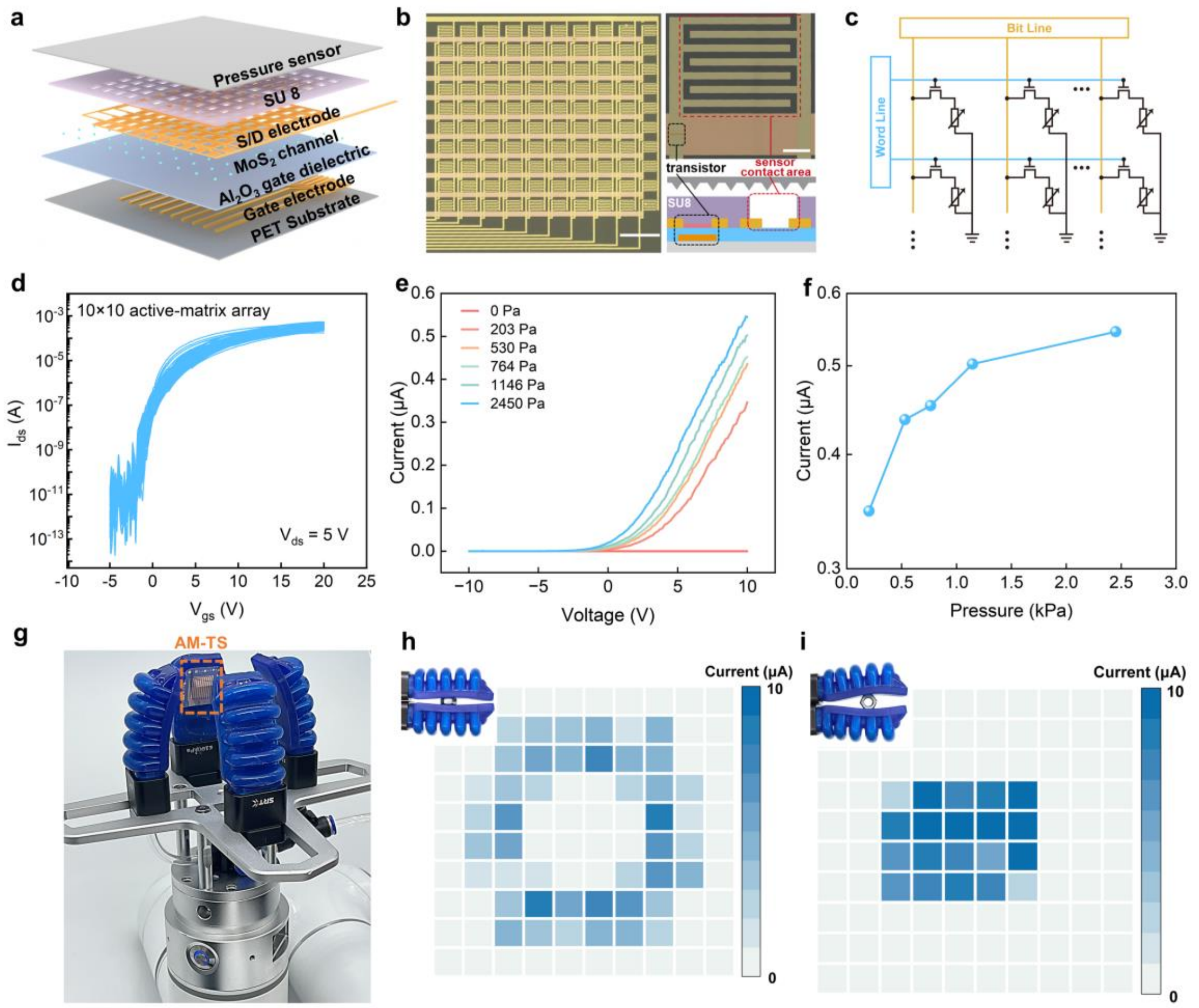

**Fig. 5 The Characteristics of the active-matrix tactile sensor array. a,** Schematic diagram of 10×10 active-matrix tactile sensor array (AM-TS). **b,** Optical image of the AM-TS with a sensing area of 1.5×1.5 cm. Scale bar, 0.3 cm. (left panel). And the magnified optical image of one pixel within the array (top). Scale bar, 200 mm. Schematic of a single pixel (bottom). **c,** Equivalent circuit diagram of the AM-TS. **d,** Transfer curves of the transfer curves of the 100 $MoS_2$ FETs in the active-matrix array. **e,** Transfer curves of a single pixel under different pressure ($V_{ds}$ = 0.1 V). **f,** Relative on current as a function of applied pressure ($V_{ds}$ = 0.1 V and $V_{gs}$ = 10 V), the data was extracted from Figure 5e. **g,** Photograph of the large area AM-TS on the finger of a soft robotic gripper system. **h-i,** On-state current distribution of AM-TS when pressing on different shapes. Inset, the photograph of an M6 Hex Nut taken from various angles within the soft robotic gripper system.

# Supporting Information for

## Wafer-scale Integration of Single-Crystalline $MoS_2$ for Flexible Electronics Enabled by Oxide Dry-transfer

**Supplementary Information:**

**Supplementary Table 3** | Comparison of gain-power consumption of reported flexible thin-film inverters.
**Supplementary Fig. 15** | Pressure sensor characterization.
**Supplementary Note 5** | The working mechanism of pressure sensor.

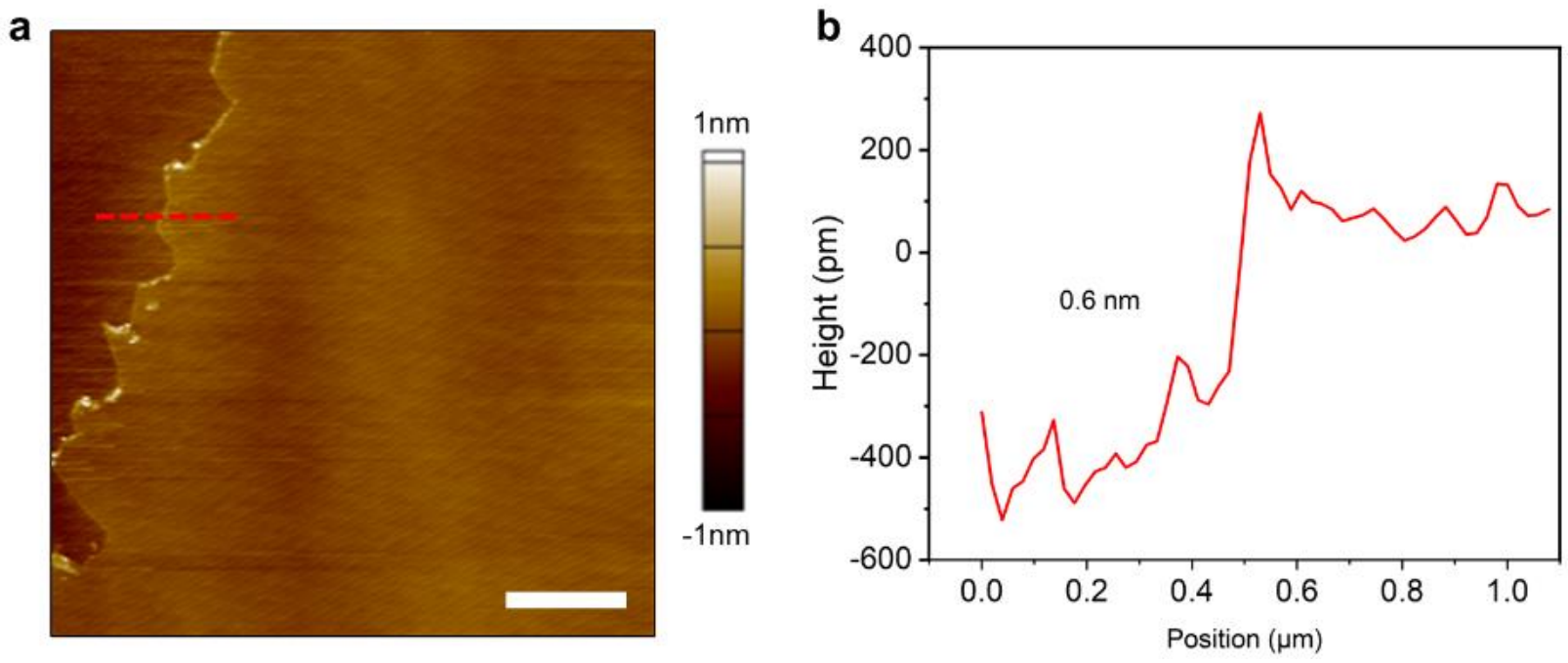

**Supplementary Fig. 1 | AFM images of $MoS_2$ film. a,** Typical AFM images of monolayer single-crystalline $MoS_2$ on sapphire, scale bar, 2 nm. **b,** Height profile of the red line shown in (a).

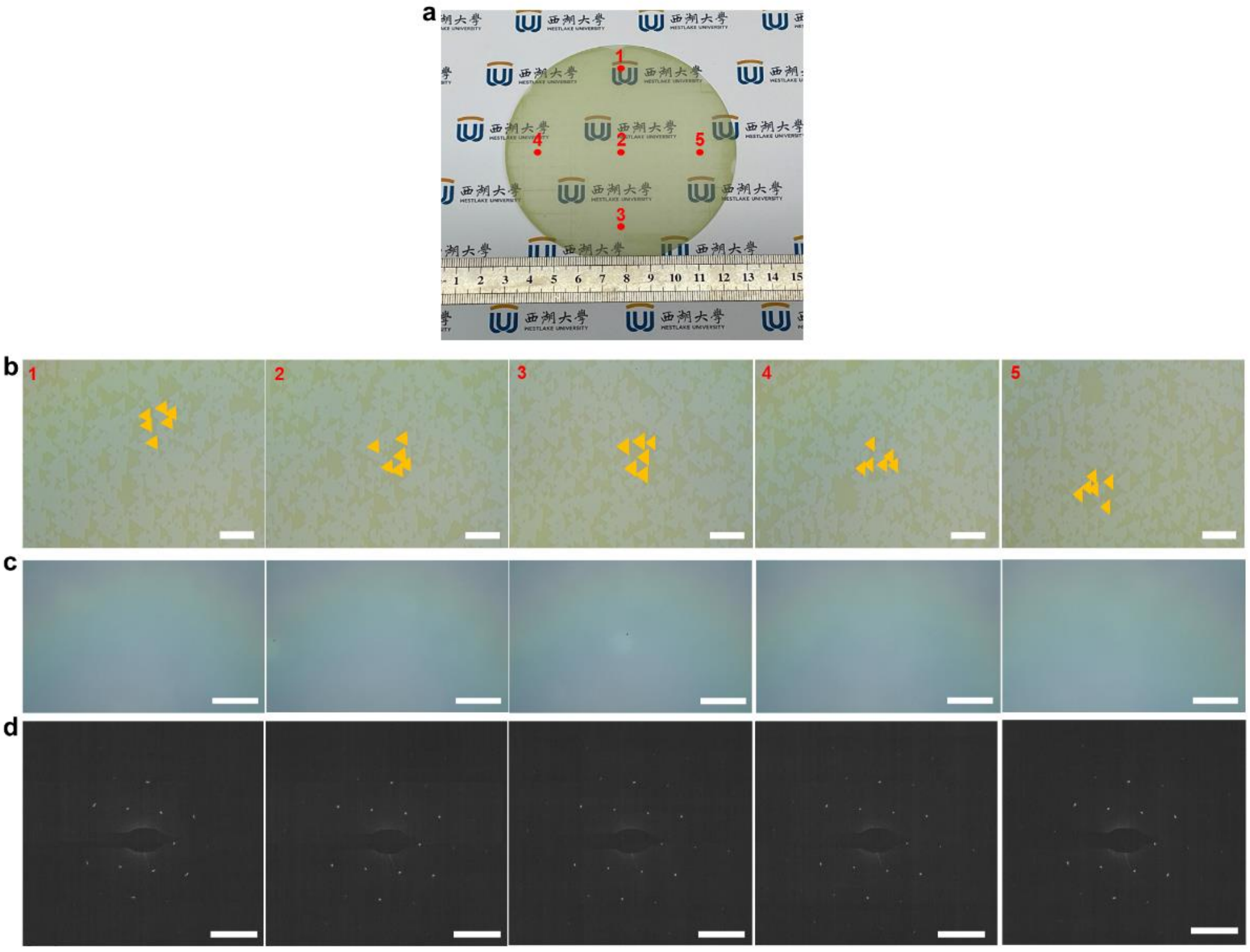

**Supplementary Fig. 2 | Basic characterization of monolayer single-crystalline $MoS_2$. a-b,** Typical optical microscopy images of $MoS_2$ crystal domains crossing wafer scale grown at 5 different positions shown in (**a**) before they merge into a continuous thin film. All the triangle flakes exhibit the same orientation. scale bar, 100 μm. The crystal domain is outlined by a yellow triangle. **c,** the optical images of continuous monolayer $MoS_2$ film. scale bar, 100 μm. **d,** Selected area electron diffraction (SAED) at five random positions. scale bar, 4 $nm^{-1}$.

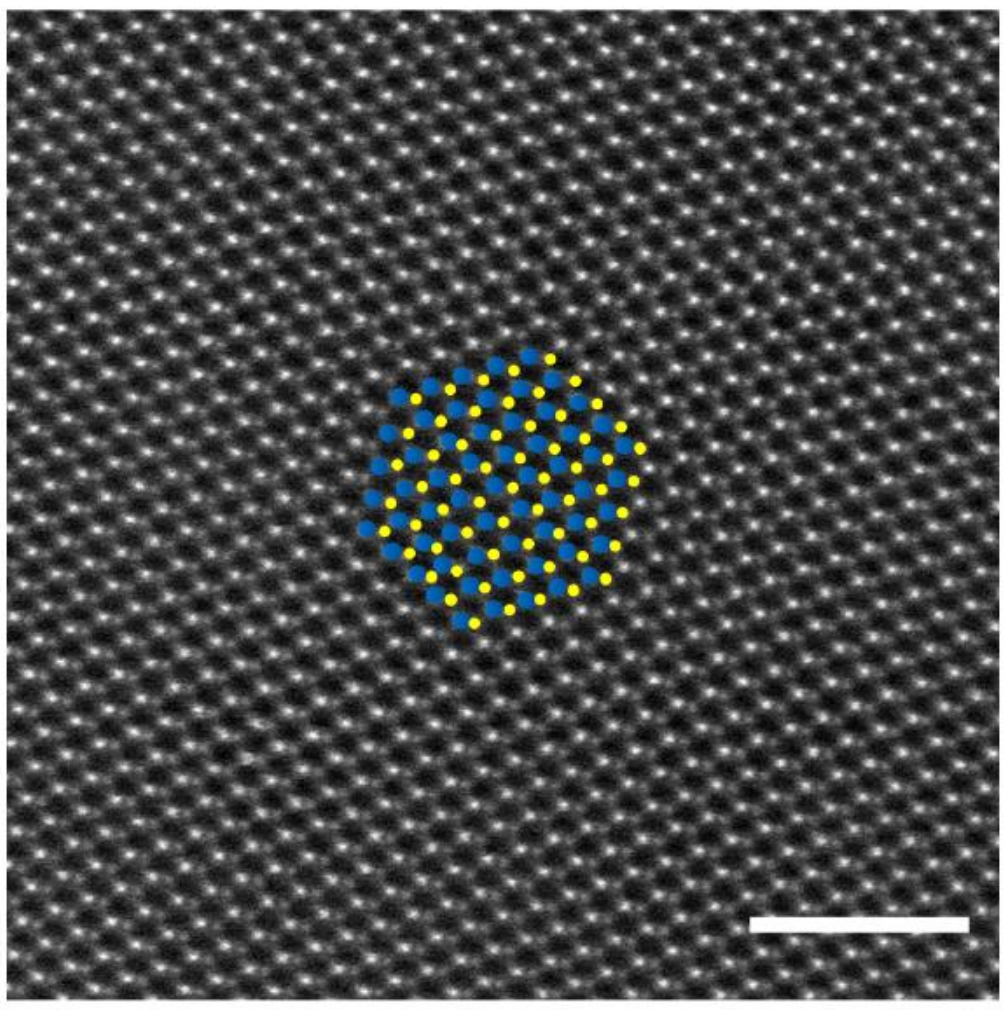

**Supplementary Fig. 3 | Atomic-resolution HAADF-STEM image of single-crystalline $MoS_2$.** The $MoS_2$ basal plane. The yellow balls marked in the figure represent sulfur atoms, and the blue balls represent molybdenum atoms. Scale bar, 2 nm.

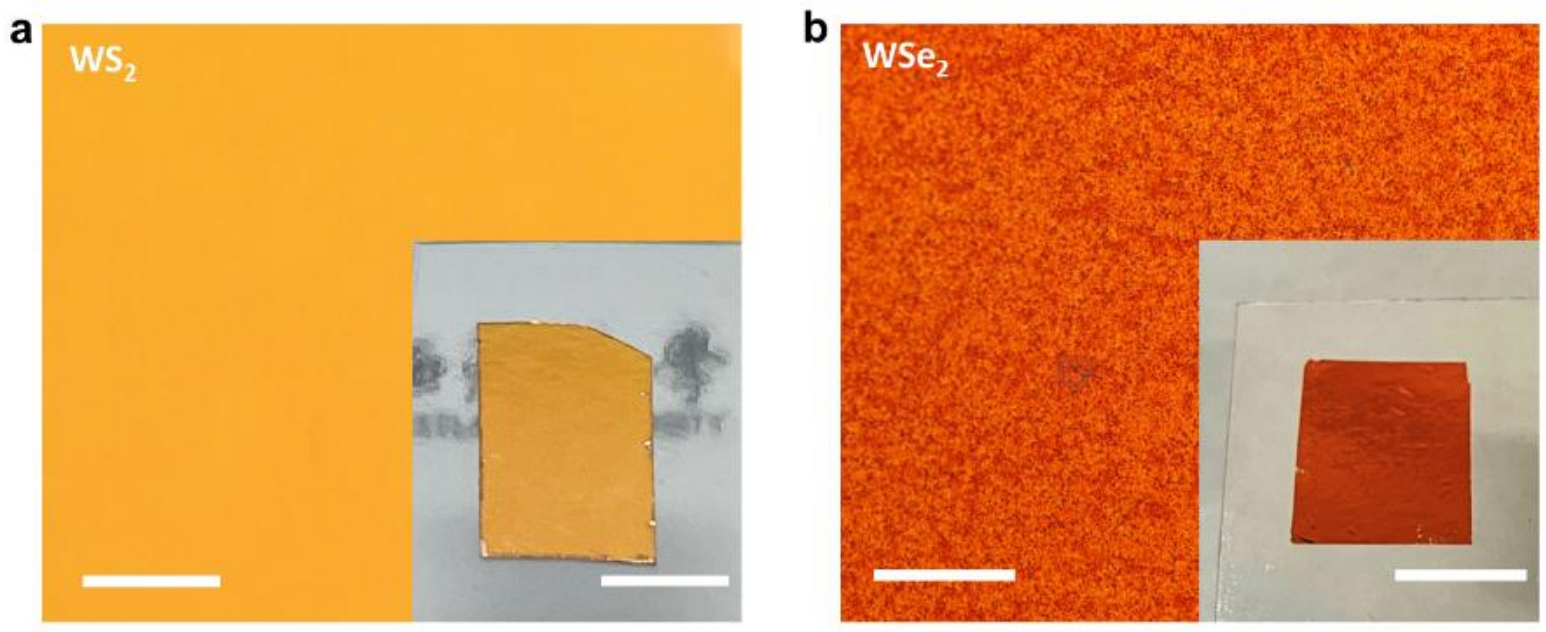

**Supplementary Fig. 4 | Material Compatibility of oxide dry-transfer method.** Optical images of $WS_2$ **(a)** and $WSe_2$ **(b)** after transfer. Scale bar, 200 μm. Their inserts are all Photographic pictures. Scale bar, 1 cm.

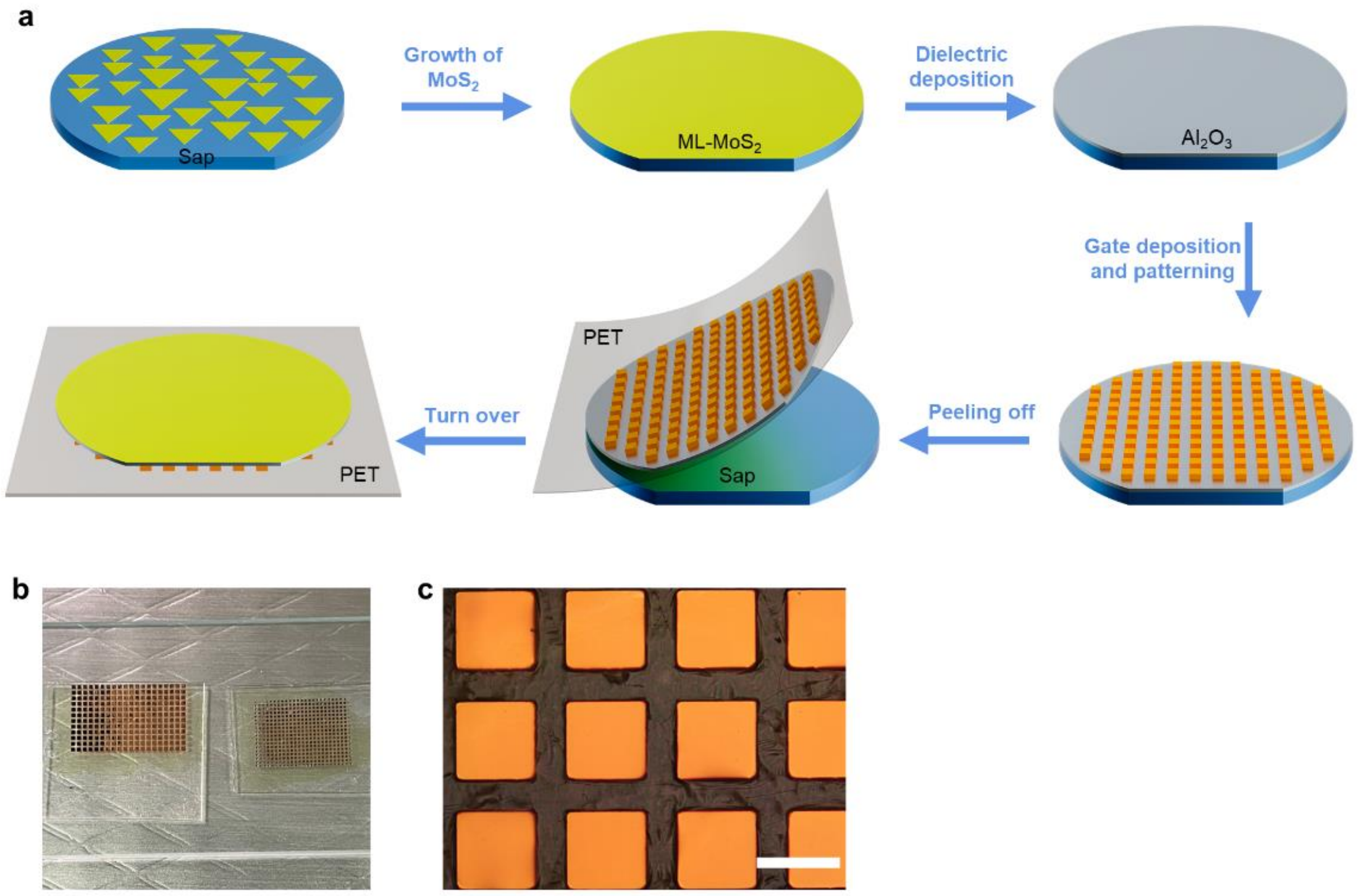

**Supplementary Fig. 5 | Oxide dry-transfer with local back gates process. a,** Schematic diagram of the $Al_2O_3$ assisted dry transfer of monolayer $MoS_2$. **b,** Photographic picture of $MoS_2$ after transfer**.** The left: Gate size 400 μm. The right: Gate size 100 μm. **c,** Optical image of $MoS_2$ after transfer.

**Supplementary Note 1 | Oxide dry-transfer with local back gates for device.**

Monolayer, single-crystalline $MoS_2$ was epitaxially synthesized on c-plane sapphire via CVD. A thin layer of $Al_2O_3$ was first deposited onto the $MoS_2$ surface using E-beam evaporation for adhesion enhancement, followed by ALD of $Al_2O_3$ as the high-κ dielectric. Subsequently, pattern the gates (Cu) through photolithography, E-beam evaporation and lift-off (It is also possible to deposit metal directly by covering a hard mask), and the entire structure was adhered to a flexible substrate, such as polyethylene terephthalate (PET), and is exfoliated from the sapphire substrate.

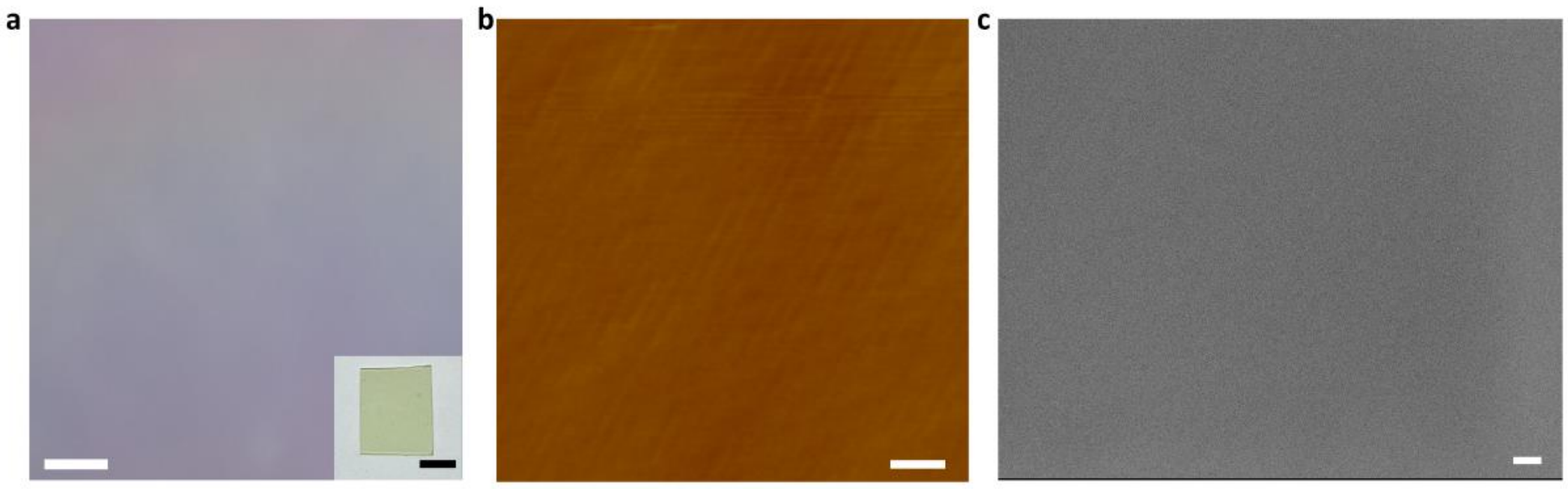

**Supplementary Fig. 6 | Cleanliness characterization of $MoS_2$ before transfer. a,** Optical images of continuous $MoS_2$ film on sapphire. Scale bar, 200 μm. Insert: photographic picture. Scale bar, 1 cm. **b,** AFM images, scale bar, 2 nm. **c,** SEM images, scale bar, 100 nm.

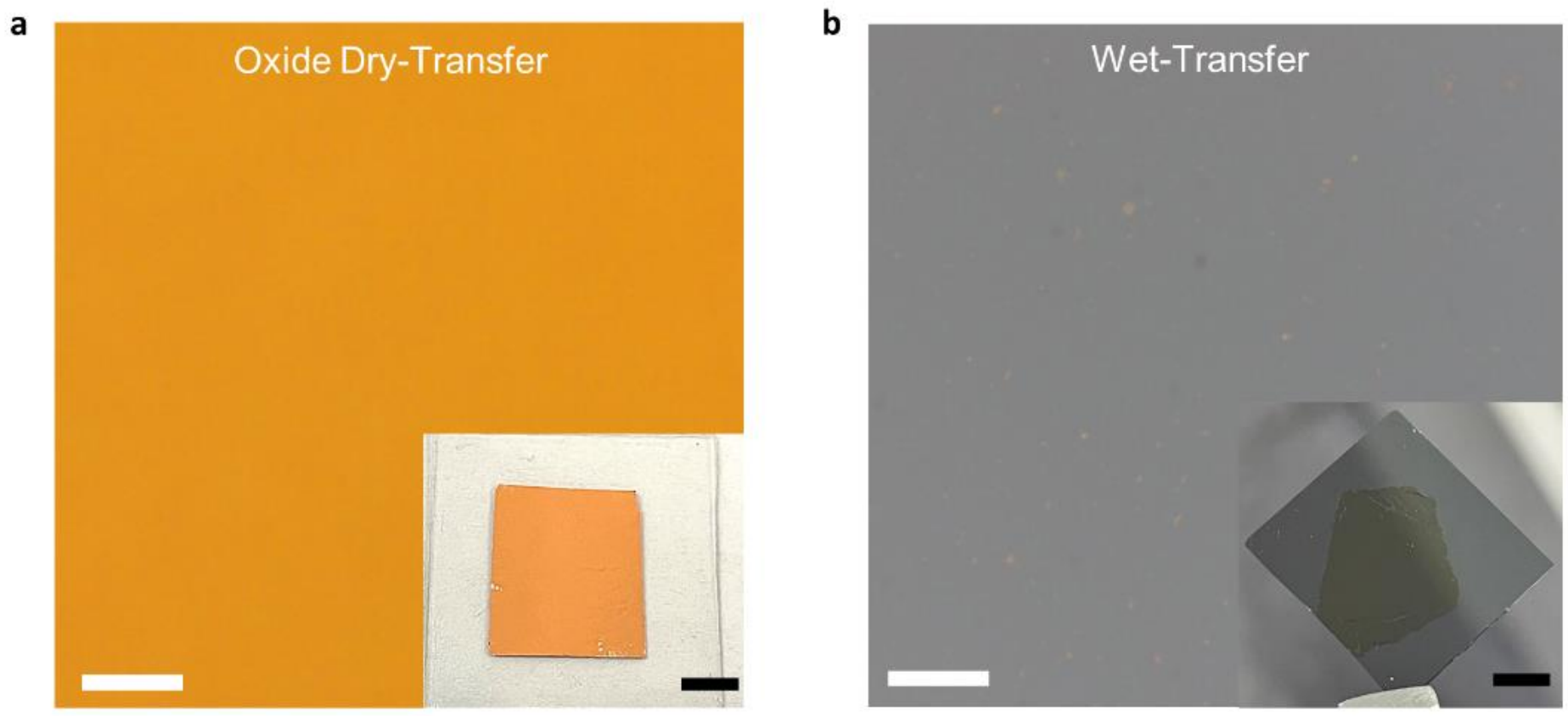

**Supplementary Fig. 7 | Cleanliness comparison of two transfer methods.** Optical micrographs of **(a)** oxide dry-transfer and **(b)**PMMA wet-transfer. Scale bar, 200 μm. Their inserts are all Photographic pictures. Scale bar, 1 cm.

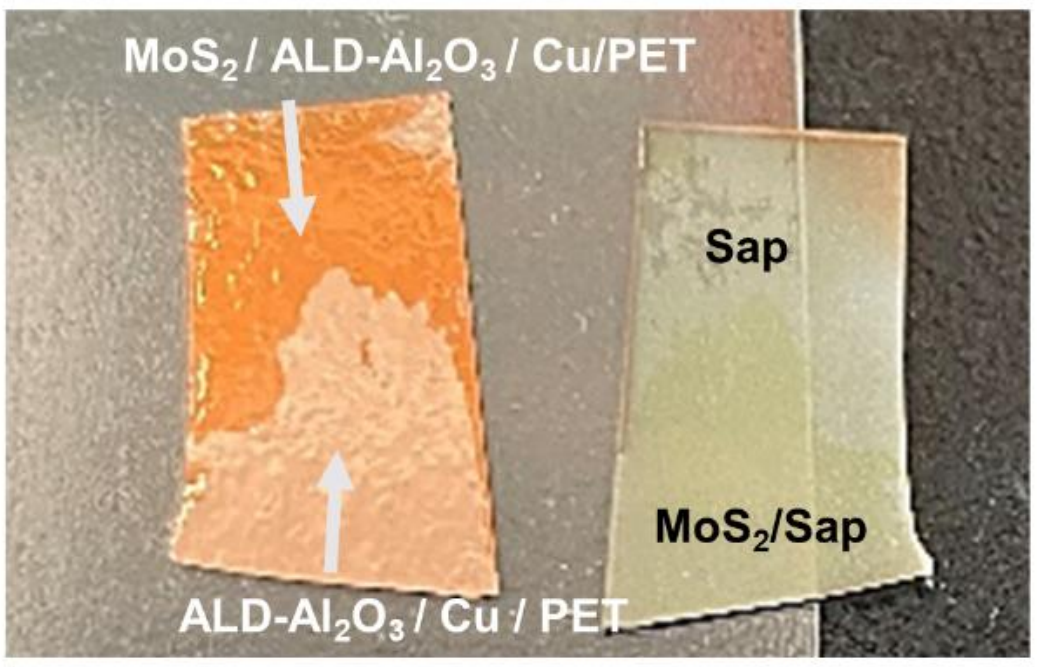

**Supplementary Fig. 8 | Oxide dry-transfer without E-beam dielectrics.** Photographic picture before and after transfer.

**Supplementary Note 2 | Effects of different oxide deposition methods on oxide dry-transfer process.**

The sample in the Supplementary Fig. 4 is a $MoS_2$ surface with 30nm $Al_2O_3$ deposited by ALD process, and then 500nm Cu deposited by E-beam, and then peeled off with PET tape, lacking the 4nm E-beam auxiliary peeling layer. We can find that the peeling effect is not stable enough, and some materials can be transferred, such as the structure of $MoS_2$/ALD-$Al_2O_3$/Cu/PET in the upper part of the figure, but some materials are not successfully transferred, such as the structure of ALD-$Al_2O_3$/Cu/PET in the lower part of the figure, which is caused by the unstable bonding force of the ALD deposition method. The difference in adhesion strength between oxide films deposited by E-beam and ALD on 2D materials can be attributed to the distinct deposition mechanisms of these two techniques.

In the case of E-beam, material is vaporized by a high-energy electron beam and then condenses onto the surface of the 2D material. The evaporated species typically possess high kinetic energy, which facilitates strong physical interactions with the substrate. This increased kinetic energy during deposition can lead to enhanced interfacial adhesion, potentially through the formation of stronger physical bonds or even localized chemical interactions between the deposited film and the substrate. This is especially true when the 2D material surface exhibits limited chemical reactivity. In contrast, ALD relies on a stepwise chemical vapor deposition process where film growth occurs through surface-controlled, self-limiting reactions. While ALD can achieve uniform and conformal thin films, the lower deposition energy and the reliance on chemical reactions between the precursor molecules and the substrate surface result in generally weaker adhesion. This is particularly evident when the 2D material surface lacks reactive sites, reducing the effectiveness of the chemical bonding process. Thus, the higher kinetic energy associated with E-beam tends to promote stronger adhesion through more robust physical interactions compared to the relatively mild, surface-controlled reactions of ALD.

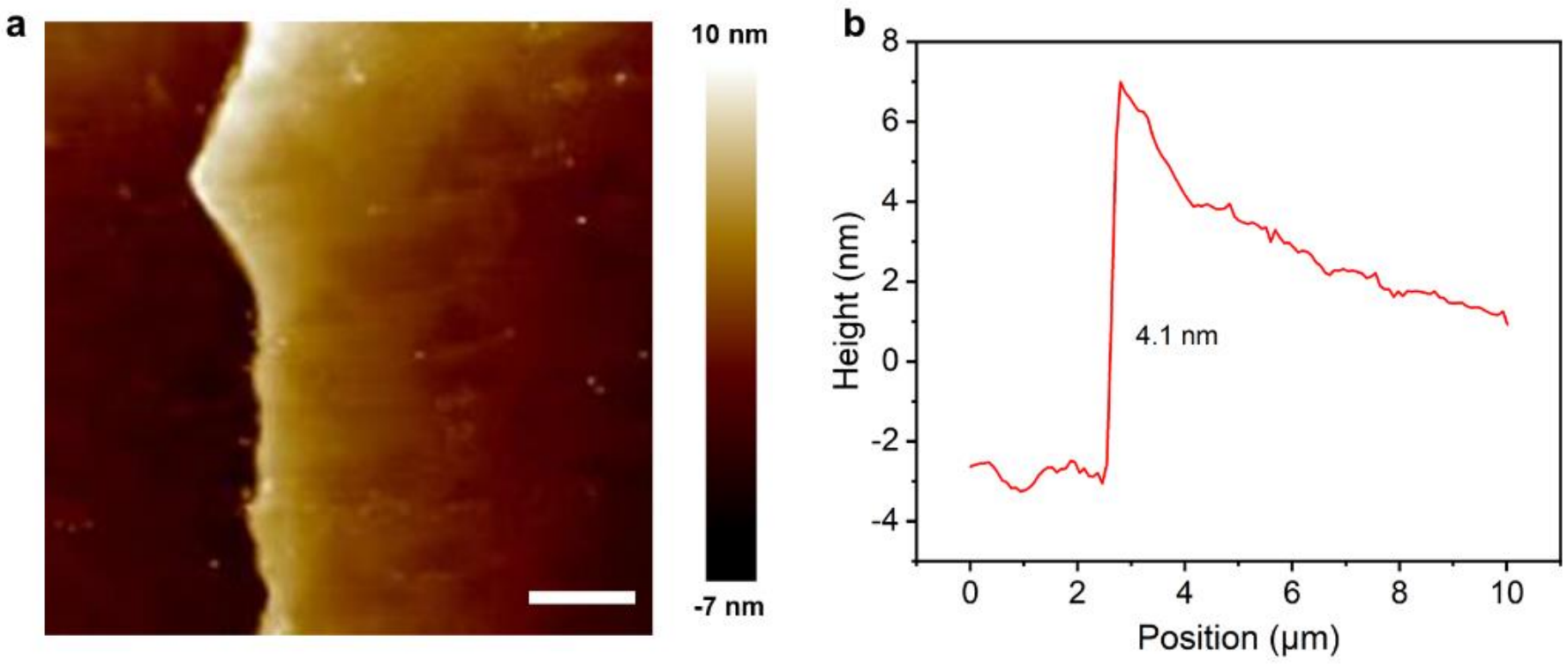

**Supplementary Fig. 9 | AFM images of E-beam-$Al_2O_3$. a,** Typical AFM images of E-beam-deposited on $MoS_2$, scale bar, 2 μm. **b,** Height profile of the red line shown in (a).

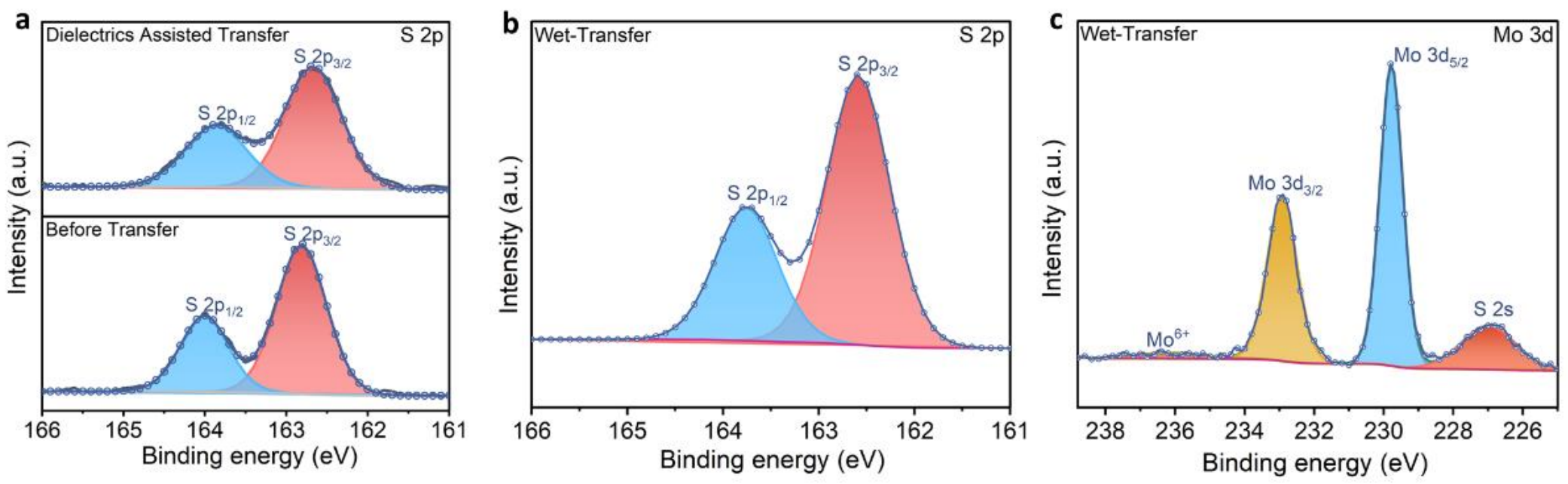

**Supplementary Fig. 10 | XPS of pristine CVD-$MoS_2$, oxide dry-transfer and wet-transfer $MoS_2$. a,** Mo 3d of $MoS_2$ before and after dielectrics assisted transfer. **b,** S 2p of $MoS_2$ wet-transfer. **c,** Mo 3d of $MoS_2$ wet-transfer.

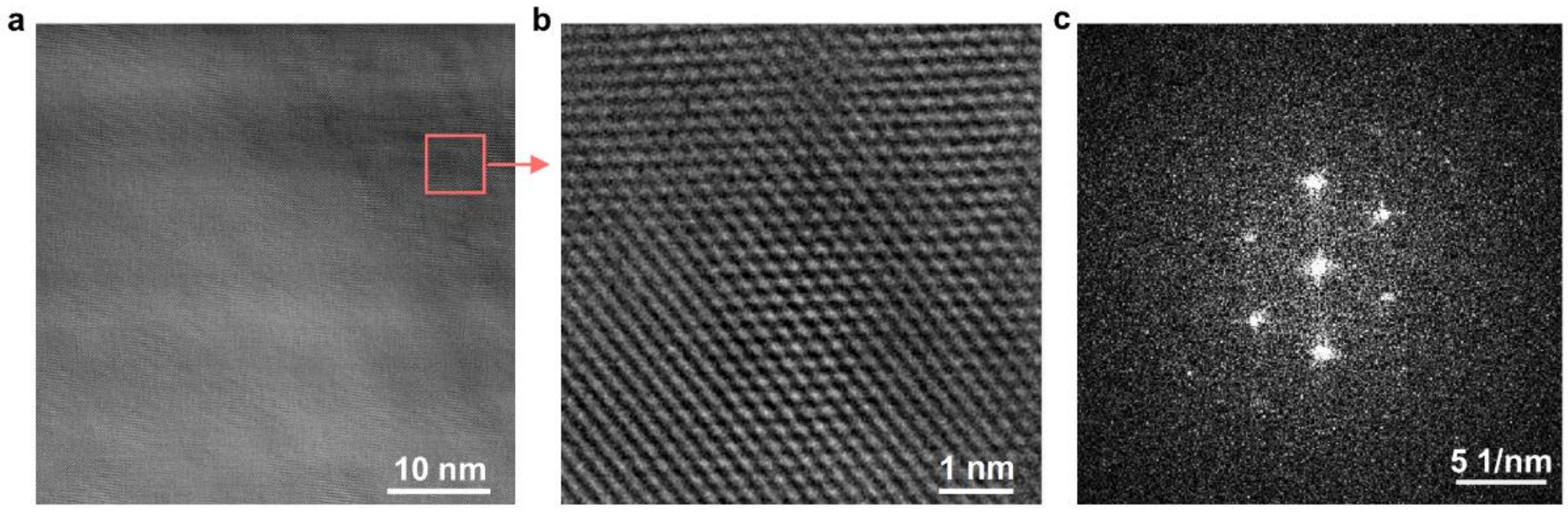

**Supplementary Fig. 11 | Crystal quality of the ALD-deposited dielectric layer. a,** HRTEM of 30nm ALD-$Al_2O_3$. **b,** The area marked by the red box in **(a)** is enlarged. **c,** According to Figure a, a set of diffraction spots are obtained by Fast Fourier Transform (FFT), which shows that the crystal orientation of the material is consistent, proving that the material has high crystallinity.

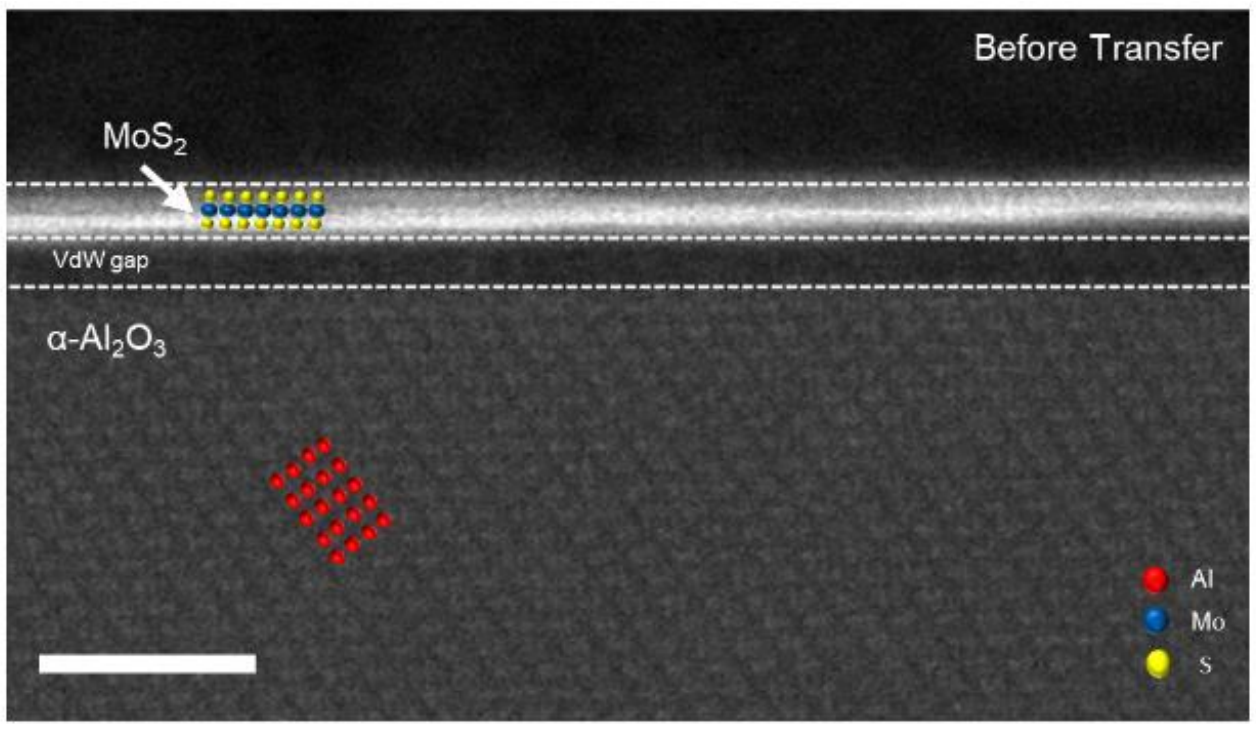

**Supplementary Fig. 12 | Cross-sectional HAADF-STEM images of the $MoS_2$ before transfer.** $MoS_2$/ $Al_2O_3$ interface along $Al_2O_3$ <$11\bar{2}0$> directions. Scale bars, 2 nm.

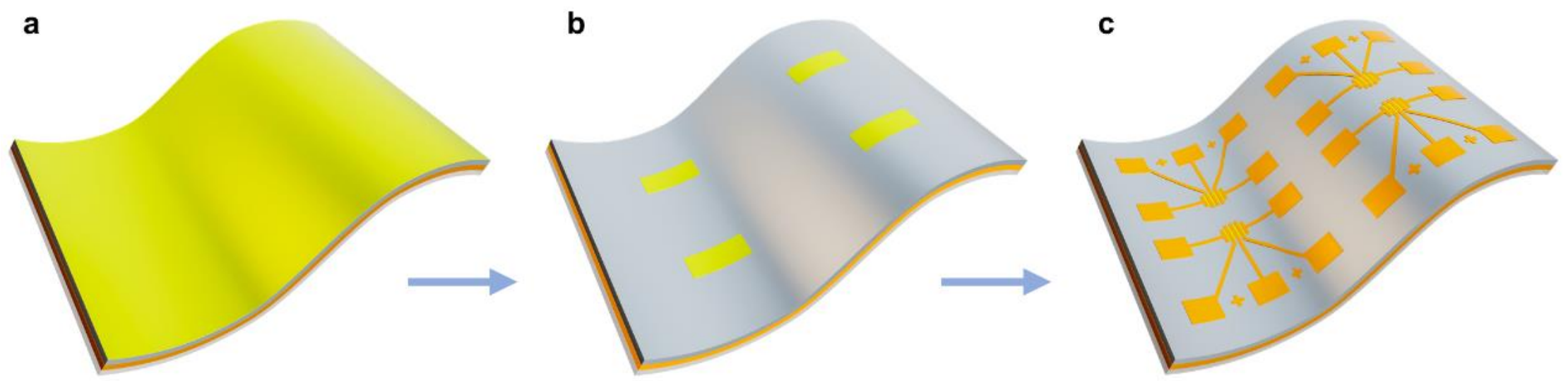

**Supplementary Fig. 13 | The process of fabricating $MoS_2$ FETs with oxide dry-transferred materials.**

**Supplementary Note 3：** As illustrated in Fig. 1a, following the complete process of oxide dry-transfer, we achieved the device structure of $MoS_2$/high-κ/Gate metal/PET substrate (Supplementary Fig. 12a). We then defined the material channel shape using photolithography (Suss-MA/BA6 Gen4, Photoresist: ARP-5350), followed by dry etching (Samco-RIE-230ip: $O_2$ 100W 2min) to remove the excess material. Subsequently, the photoresist mask was removed by immersing the structure in acetone for 10 minutes, isopropanol for 5 minutes, and anhydrous ethanol for 1 minute, resulting in the structure depicted in Supplementary Fig. 12b. Finally, after a second photolithography step to define the source/drain (S/D) electrode pattern, we deposited 20 nm of Bi and 30 nm of Au via electron beam evaporation. The final structure was obtained by lifting off, completing the back-gate FETs array as shown in Supplementary Fig. 12c. All electrical measurements were carried out by a Keithley 4200A-SCS semiconductor parameter analyzer in a closed-cycle cryogenic probe station with base pressure $10^{-6}$ mba at room temperature.

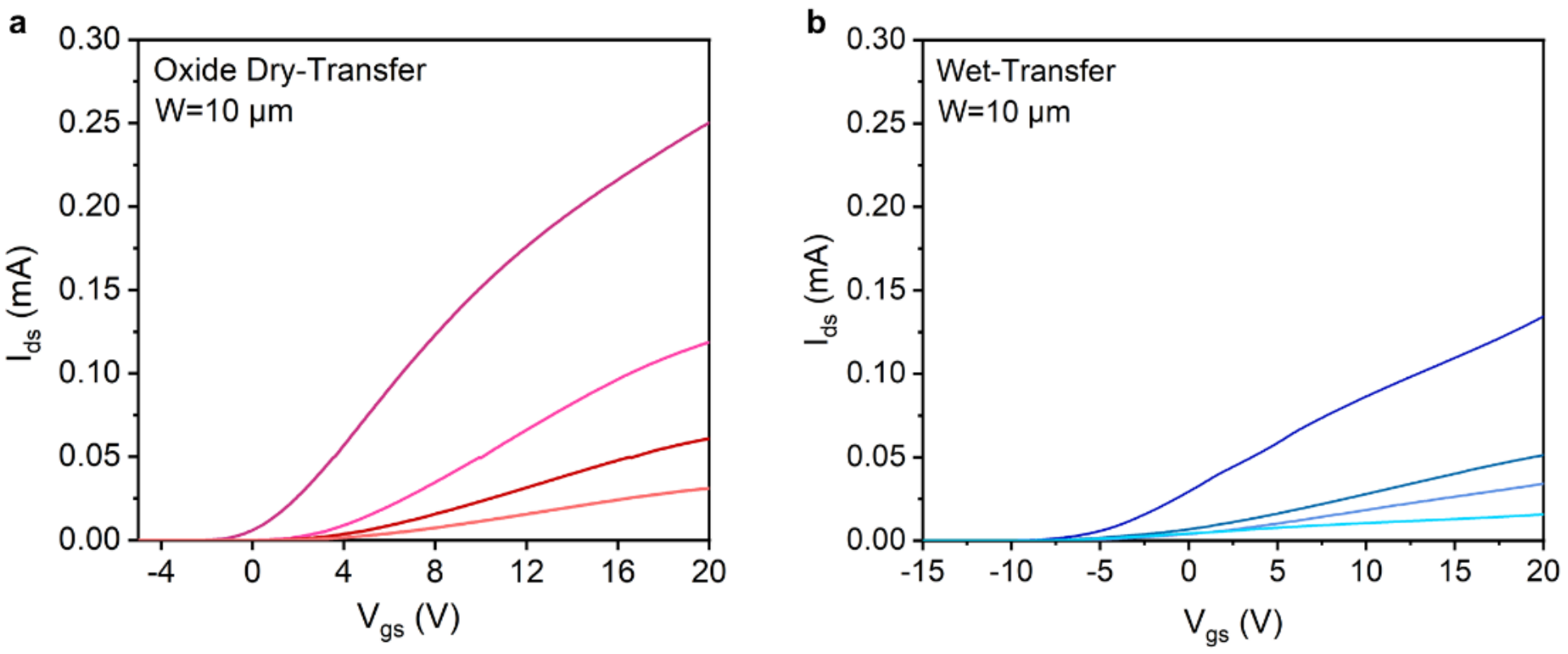

**Supplementary Fig. 14 | Transfer characteristics of Bi-contacted $MoS_2$ FETs with various channel length ($L_{ch}$) at $V_{ds} = 1$ V for the TLM study.** FETs fabricated from materials transferred by oxide dry-transfer **(a)** and wet-transfer **(b)**. The $L_{ch}$ are 2 μm, 6 μm, 10 μm and 20 μm respectively.

**Supplementary Note 4: Field effect mobility calculation**

The mobility was calculated by the formula

$$\mu = \frac{L}{W} \cdot \frac{1}{C_{ox}} \cdot \frac{1}{V_{ds}} \cdot \frac{\partial I_{ds}}{\partial V_{gs}}$$

where $L$ and $W$ are the channel length and width and $C_{ox}$ is the capacitance of $Al_2O_3$. $Cox =$1.1×10$^{-7}$ F/cm$^2$ by measured directly. $V_{ds}$, $I_{ds}$ and $V_{gs}$ is calculated from the transfer curve of the device.

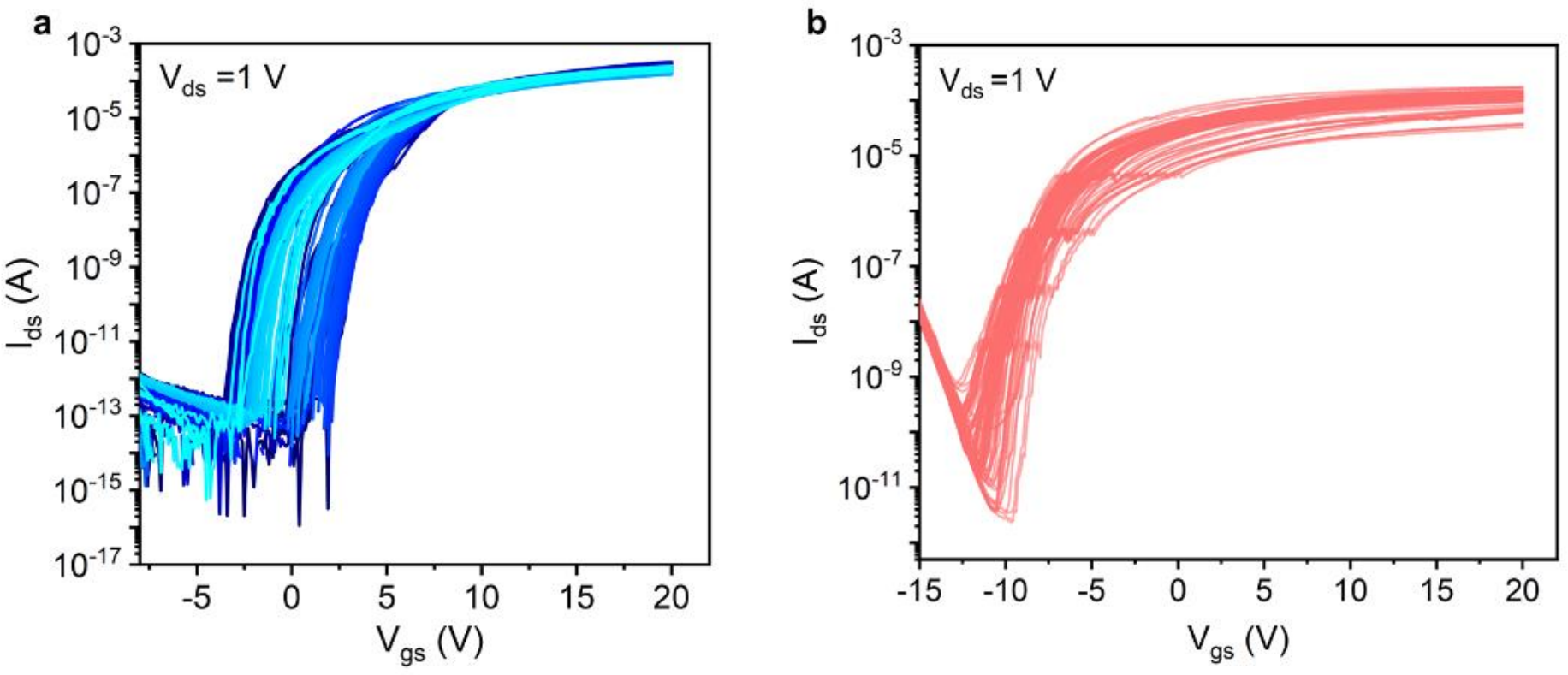

**Supplementary Fig. 15 | 100 randomly picked FETs performance.** Transfer curve of oxide dry-transfer **(a)** and wet-transfer **(b)** $MoS_2$ FET devices, with $V_{ds}$=1V.

**Supplementary Table 1.**

**Comparison of $MoS_2$ electrical performance on flexible substrates.**

| Synthesis | Layer | $T_{ox}$ (nm) | Mobility ($cm^2V^{-1}s^{-1}$) | On/off | SS (mV $dec^{-1}$) | Refs. |
|---|---|---|---|---|---|---|
| CVD | 1L-$MoS_2$ | 35nm $Al_2O_3$ | 117 | $10^{12}$ | 68 | This work |
| CVD | 1L-$MoS_2$ | 35nm $Al_2O_3$ | 55 | $10^{10}$ | 167 | Ref.[1] |
| CVD | 1L-$MoS_2$ | 50nm $Al_2O_3$ | 16.2 | $10^6$ | / | Ref.[2] |
| CVD | 1L-$MoS_2$ | $SiO_2$ | 12.5 | $10^5$ | / | Ref.[3] |
| CVD | 1L-$MoS_2$ | 10nm $HfO_2$ | 70 | $10^9$ | 70 | Ref.[4] |
| CVD | 1L-$MoS_2$ | 35nm $Al_2O_3$ | 27 | $10^6$ | 730 | Ref.[5] |
| CVD | 1L-$MoS_2$ | 30nm $Al_2O_3$ | 9.1 | $10^8$ | 370 | Ref.[6] |
| CVD | 1L-$MoS_2$ | 30nm $HfO_2$ | 13.9 | $10^5$ | / | Ref.[7] |
| CVD | 1L-$MoS_2$ | 30nm $HfO_2$ | 22 | $10^5$ | / | Ref.[8] |
| CVD | 1L-$MoS_2$ | 35nm $Al_2O_3$ | 18.9 | $10^7$ | 518 | Ref.[9] |
| MOCVD | 1L-$MoS_2$ | 30nm $Al_2O_3$ | 1.8 | $10^8$ | 300 | Ref.[10] |
| MOCVD | 1L-$MoS_2$ | 50nm $Al_2O_3$ | 20 | $10^6$ | / | Ref.[11] |
| Exfoliated | Multilayer $MoS_2$ | 300nm SiO2 | 42(-223k) | ~ $10^6$ | ~ 100 | Ref.[12] |
| Exfoliated | 1L-$MoS_2$ | 280nm SiO2 | 12 | $10^6$ | / | Ref.[13] |
| Exfoliated | Few-layer $MoS_2$ | 30nm $Al_2O_3$ | 20 | $10^9$ | ~160 | Ref.[14] |
| Exfoliated | Multilayer $MoS_2$ | 400nm $SiO_2$ | 4.7 | $10^4$ | / | Ref.[15] |
| Exfoliated | Multilayer $MoS_2$ | 25nm $Al_2O_3$ | 19 | $10^6$ | 250 | Ref.[16] |
| Exfoliated | Multilayer $MoS_2$ | 25nm $Al_2O_3$/$HfO_2$ | 30 | $10^7$ | 82 | Ref.[17] |
| Exfoliated | Multilayer $MoS_2$ | 80nm $Al_2O_3$ | 18.12 | $10^5$ | / | Ref.[18] |
| ALD | Multilayer $MoS_2$ | 30nm $Al_2O_3$ | 32 | $10^7$ | 80 | Ref.[19] |

**Supplementary Table 2.**

**Comparison of $MoS_2$ electrical performance on rigid substrates.**

| Synthesis | Layer | $T_{ox}$ (nm) | Mobility ($cm^2V^{-1}s^{-1}$) | On/off | SS (mV $dec^{-1}$) | Ref. |
|---|---|---|---|---|---|---|
| CVD | 1L-$MoS_2$ | 35nm $Al_2O_3$ | 117 | $10^{12}$ | 68 | This work |
| CVD | 1L-$MoS_2$ | 12nm HfLaO | 36 | $2.3\times10^{11}$ | / | Ref.[20] |
| CVD | 1L-$MoS_2$ | 300nm $SiO_2$ | 21 | $2\times10^{7}$ | / | Ref.[21] |
| CVD | 1L-$MoS_2$ | 15nm $HfO_2$ | 118 | $10^{8}$ | / | Ref.[22] |
| CVD | 1L-$MoS_2$ | 300nm $SiO_2$ | 23 | $10^{6}$ | / | Ref.[23] |
| CVD | 1L-$MoS_2$ | 15nm $HfO_2$ | 131.6 | $10^{7}$ | 200 | Ref.[24] |
| CVD | 1L-$MoS_2$ | $SiO_2$ | 19.2 | $3.2\times10^{9}$ | / | Ref.[25] |
| CVD | 1L-$MoS_2$ | 4nm $HfO_2$ | 1.48 | $10^{7}$ | / | Ref.[26] |
| CVD | 1L-$MoS_2$ | 30nm $HfO_2$ | 102.6 | $10^{9}$ | 140 | Ref.[27] |
| CVD | 1L-$MoS_2$ | 20nm $Al_2O_3$ | 70 | $10^{9}$ | 200 | Ref.[28] |
| CVD | 1L-$MoS_2$ | 3nm $HfO_2$ | 34 | $10^{7}$ | 60 | Ref.[29] |
| CVD | 1L-$MoS_2$ | 300nm $SiO_2$ | 40 | $10^{6}$ | / | Ref.[30] |
| CVD | 1L-$MoS_2$ | 300nm $SiO_2$ | 82 | $2\times10^{10}$ | / | Ref.[31] |
| CVD | 1L-$MoS_2$ | 20nm $Al_2O_3$ | 120 | $7.3\times10^{7}$ | / | Ref.[32] |
| MOCVD | 1L-$MoS_2$ | 30nm $Al_2O_3$ | 45 | $10^{7}$ | 485 | Ref.[33] |
| MOCVD | 1L-$MoS_2$ | 100nm $SiO_2$ | 21 | $10^{7}$ | / | Ref.[34] |
| Exfoliated | 1L-$MoS_2$ | 50nm $SiO_2$ | 26 | $10^{5}$ | / | Ref.[35] |
| Exfoliated | 1L-$MoS_2$ | 100nm $SiO_2$ | 42 | $1.7\times10^{10}$ | / | Ref.[36] |
| Exfoliated | 1L-$MoS_2$ | 300nm $SiO_2$ | 30 | $10^{7}$ | 74 | Ref.[37] |

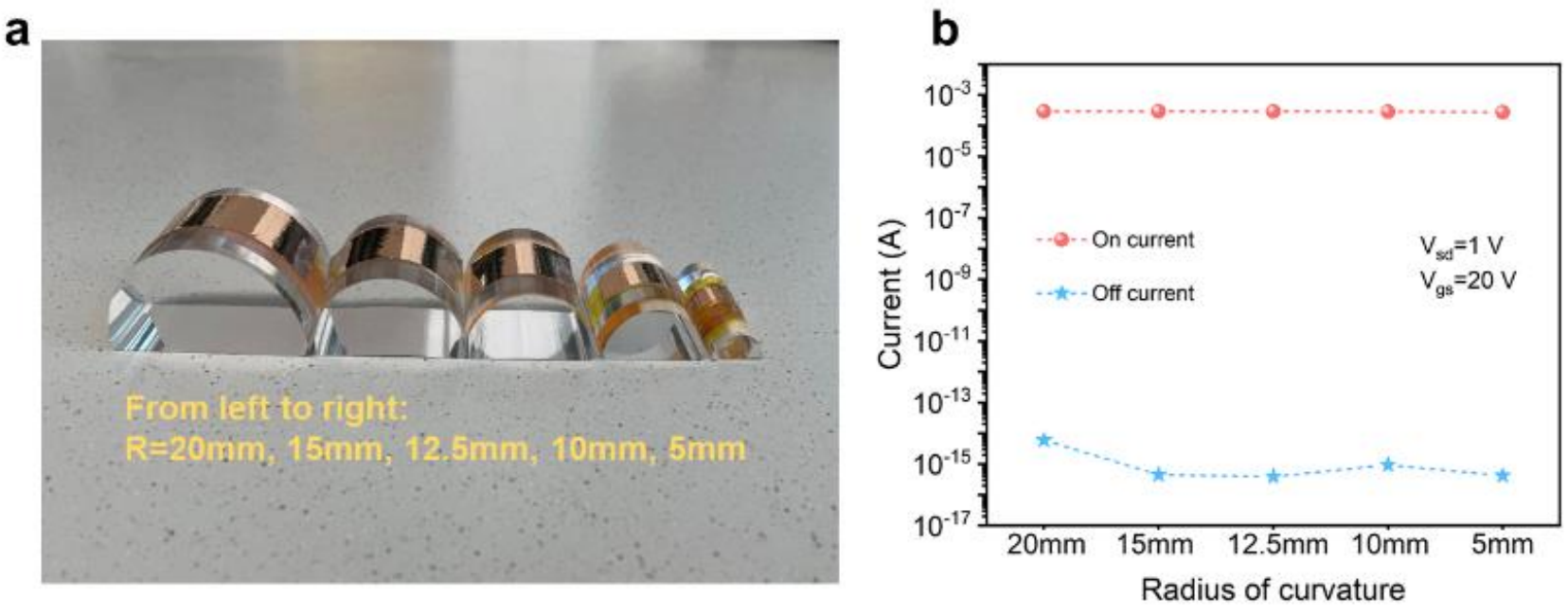

**Supplementary Fig. 16 | Bending test of flexible $MoS_2$ FETs. a,** Photographic picture of the test mold, the curvature increases from left to right, we glued the flexible device to its arc surface for subsequent electrical testing. **b,** Comparison of device on-state current/off-state current under different curvature radii.

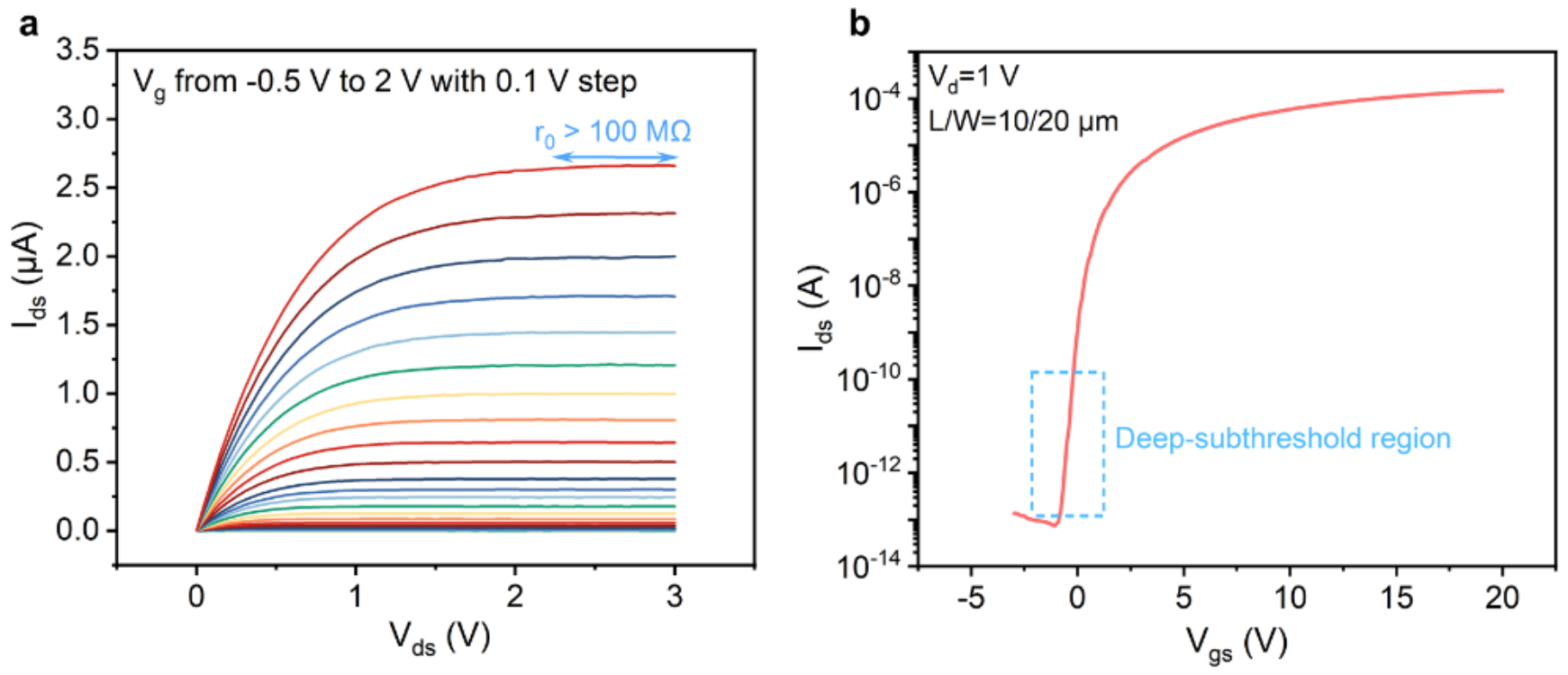

**Supplementary Fig. 17 | Characterizations of the unit TFT in an inverter. a.** Output curves of load FET with L=10 μm and W=20 μm. **b.** Corresponding transfer curves of the device.

**Supplementary Table 3.**

**Comparison of gain-power consumption of reported flexible thin-film inverters**

| Material Type | Material | Inverter structure | Gain (V/V) | Power consumption (W/μm) | Refs. |
|---|---|---|---|---|---|
| TMDCs | $MoS_2$ | NMOS | 216 | $1.2\times10^{-12}$ | This work |
| TMDCs | $MoS_2$ | NMOS | 120 | $1.03\times10^{-11}$ | Ref.[4] |
| TMDCs | $MoS_2$ | NMOS | 43 | $4.0\times10^{-10}$ | Ref.[1] |
| TMDCs | $MoS_2$ | CMOS | 8.2 | $5.8\times10^{-11}$ | Ref.[38] |
| TMDCs | $MoS_2+WSe_2$ | CMOS | 100 | $6.0\times10^{-8}$ | Ref.[39] |
| Oxide | ZnO | NMOS | 40 | $1.4\times10^{-6}$ | Ref.[40] |
| Oxide | IGZO | NMOS | 2.45 | $7.9\times10^{-6}$ | Ref.[41] |
| Oxide | IGZO | NMOS | 8.61 | $1.8\times10^{-5}$ | Ref.[42] |
| Organic | N1100 | CMOS | 90 | $1.0\times10^{-12}$ | Ref.[43] |
| Organic | DNTT | CMOS | 56 | $1.4\times10^{-8}$ | Ref.[44] |
| Organic | PTAA/Acene based-diimide | CMOS | 29 | $4.0\times10^{-10}$ | Ref.[45] |
| Organic | P(g42T-T)/ BBL | CMOS | 26 | $1.35\times10^{-9}$ | Ref.[46] |
| Hybrid | IGZO/DPP-G2T | CMOS | 95 | $6.6\times10^{-11}$ | Ref.[47] |
| Hybrid | ITO/Chitosan | NMOS | 19 | $2.0\times10^{-8}$ | Ref.[48] |
| Hybrid | IGZO/CNT | CMOS | 45 | $6.9\times10^{-12}$ | Ref.[49] |
| Hybrid | IGZO/CNT | CMOS | 20 | $9.5\times10^{-12}$ | Ref.[50] |
| Hybrid | ZnO/P3HT | CMOS | 18 | $6.0\times10^{-7}$ | Ref.[51] |
| Hybrid | $MoS_2$/Si | CMOS | 16 | $3.2\times10^{-11}$ | Ref.[52] |
| CNT | | PMOS | 57 | $1.0\times10^{-9}$ | Ref.[53] |
| CNT | | PMOS | 2.5 | $7.4\times10^{-12}$ | Ref.[54] |
| CNT | | CMOS | 58 | $3.2\times10^{-6}$ | Ref.[55] |
| CNT | | CMOS | 45 | $9.0\times10^{-12}$ | Ref.[56] |
| CNT | | CMOS | 33 | $1.23\times10^{-8}$ | Ref.[57] |

| Material Type | Material | Inverter structure | Gain (V/V) | Power consumption (W/μm) | Ref. |
|---|---|---|---|---|---|
| CNT | | CMOS | 30 | $1.67\times10^{-10}$ | Ref.[58] |
| CNT | | CMOS | 23 | $9.0\times10^{-10}$ | Ref.[59] |
| CNT | | CMOS | 22 | $3.0\times10^{-11}$ | Ref.[60] |
| CNT | | CMOS | 10 | $8.6\times10^{-12}$ | Ref.[61] |
| CNT | | CMOS | 3.29 | $4.8\times10^{-12}$ | Ref.[62] |
| CNT | | CMOS | 3.1 | $3.05\times10^{-11}$ | Ref.[63] |
| a-Si | | NMOS | 1.0 | $4.0\times10^{-7}$ | Ref.[64] |
| LTPS | | CMOS | 38 | $5.6\times10^{-9}$ | Ref.[65] |
| LTPS | | CMOS | 20 | $8.3\times10^{-9}$ | Ref.[66] |

**Notes:**

1. As some of the referenced studies report power consumption data only at specific $V_{dd}$ voltages, all gain values are calculated based on the operating voltage corresponding to the reported power consumption. When multiple sets of power consumption data with different $V_{dd}$ values are available, the data corresponding to the highest gain is selected.
2. Due to the normalization of power consumption, there are inconsistencies in the N-channel and P-channel widths of certain CMOS inverters. To ensure uniformity in the analysis, the widest channel width is selected for normalization.

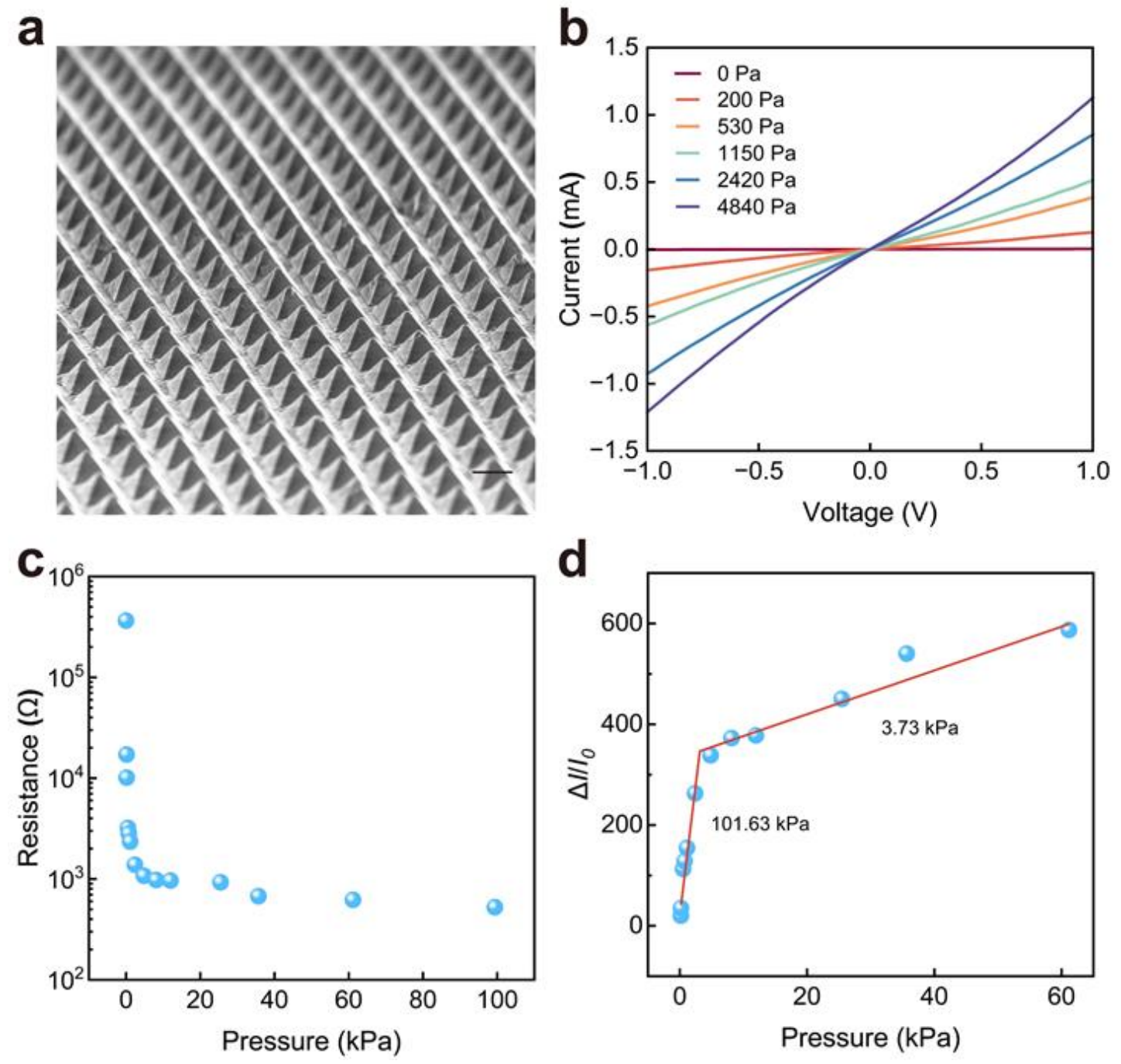

**Supplementary Fig. 18 | Pressure sensor characterization. a,** SEM image of the micro-pyramidal CNTs/PDMS resistive film. **b,** I-V curves of the CNTs/PDMS piezoresistive sensor under various pressure conditions (0-5 kPa). The initial resistance of the pressure sensor was 365 kΩ without applied pressure and decreased to 1.08 kΩ when a pressure of approximately 4840 Pa was applied. **c,** The change in resistance under various pressure conditions. **d,** The change of current response as a function of the applied pressure to extract the sensitivity of the CNTs/PDMS piezoresistive sensor. The pressure sensitivity S can be defined as S = δ(ΔI/I0)/δp, where p represents the applied pressure, I and I0 denote the current with and without applied pressure, respectively. It shows sensitivities of 101.63 and 3.73 kPa−1 in low (<3 kPa) and high (>3 kPa) pressure regimes, respectively.

**Supplementary Note 5:**

The working mechanism of a piezoresistive sensor comprised of a micro-pyramidal CNTs/PDMS film relies on the alterations induced by pressure in the conductive network established by the CNTs, facilitating the detection of mechanical stimuli. When the pressure was applied, the CNTs undergo deformation, thus modifying the interstitial spacing between them and subsequently altering the resistance of the composite material.